\documentclass[%
 reprint,
 amsmath,amssymb,
 aps,
floatfix
]{revtex4-2}

\usepackage{graphicx}
\usepackage{dcolumn}
\usepackage{bm}

\usepackage{color}
\usepackage{hyperref}
\usepackage{cleveref}
\usepackage[normalem]{ulem}
\renewcommand{\l}{\left}
\renewcommand{\r}{\right}

\begin{document}


\title{General relativistic treatment of $f$-mode oscillations of hyperonic stars}

\author{Bikram Keshari Pradhan}
\email{bikramp@iucaa.in}
\author{Debarati Chatterjee}%
 \email{debarati@iucaa.in}
\affiliation{Inter-University Centre for Astronomy and Astrophysics, \\
Pune University Campus,\\
Pune - 411007, India
}%
\author{Michael Lanoye}
\author{Prashanth Jaikumar}%
\affiliation{Department of Physics and Astronomy,\\
California State University Long Beach, Long Beach, California 90840 U.S.A.
}%

\date{\today}

\begin{abstract}
We present a systematic study of $f$-mode oscillations in neutron stars containing hyperons, extending recent results obtained within the Cowling approximation to linearized General Relativity. Employing a relativistic mean field model, we find that the Cowling approximation can overestimate the quadrupolar $f$-mode frequency of neutron stars by up to 30\% compared to the frequency obtained in the linearized general relativistic formalism. Imposing current astrophysical constraints, we derive updated empirical relations for gravitational wave asteroseismology. The frequency and damping time of quadrupole $f$-mode oscillations of hyperonic stars are found to be in the range of 1.47 - 2.45kHz and 0.13 - 0.51 sec respectively. Our correlation studies demonstrate that among the various parameters of the nucleonic and hyperonic sectors of the model, the nucleon effective mass shows the strongest correlation with mode characteristics and neutron star observables. Estimates for the detectability of $f$-modes in a transient burst of gravitational waves from isolated hyperonic stars is also provided.

\end{abstract}

                              
\maketitle


\section{Introduction}  
\label{sec:intro} 

Neutron Stars (NS) are natural laboratories to probe the behaviour of matter under extreme conditions, such as ultra-high densities, rapid rotation or ultrastrong magnetic fields~\cite{Vidana2020,Lattimer2004,Lattimer2007}. With the interior composition of the NS core uncertain, it is conjectured that strangeness in the form of hyperons, meson condensates or even deconfined quark matter may appear at such high densities, which can affect several NS observable properties. For example, the appearance of hyperons can affect NS maximum mass, radius, cooling or gravitational wave (GW) emission from unstable quasi-normal modes~\cite{VidanaEPJA}, and one can then look for the signatures of such exotic matter in NS observables.


A good theoretical model of NS should be able to explain basic NS astrophysical observables, such as its mass or radius. In order to connect the NS internal composition with these global properties, one requires an Equation of State (EoS). Various EoS models exist, that employ {\it ab-initio} many-body methods or phenomenological theories, in order to extrapolate baryon-baryon interaction to densities or neutron-proton asymmetries relevant for describing NS matter. Among the different EoS models, one class of realistic phenomenological models is based on the Relativistic Mean Field (RMF) approach, which is a particular self-consistent approximation to in-medium nuclear many-body forces and contains density dependent parameters that are fit to nuclear experimental observables ~\cite{chen2014,Hornick}. In this work, we adopt one such RMF model as a representative of this class of EoS, and call it simply ``the RMF model". 
\\

With the current generation of space-based and ground-based telescopes, neutron stars are observed at multiple wavelengths of the electromagnetic spectrum, from radio to X-rays to gamma-rays. For neutron stars in a binary, post-Keplerian effects allow the component masses to be determined to high accuracy ~\cite{Demorest2010,Antoniadis2013,Cromartie,Fonsecate,riley2021}. Radius measurements that rely solely on thermal emission from the NS surface suffer from several uncertainties, and cannot be determined with high precision. However, the recently launched NICER mission~\cite{NICER} (Neutron Star Interior Composition Explorer) has improved radius determination by employing novel techniques to study pulse modulation profiles, which enables upto 5\% accuracy in the determination of the radius ~\cite{Miller2019,Riley2019}.
\\

In addition to electromagnetic emission, neutron stars can also act as sources of gravitational waves (GW). Any non-axisymmetric perturbation or a merger of neutron stars in a binary can produce copious amounts of GWs. In the case of mergers, the tidal deformation of a component NS under the strong gravitational force of the other can constrain the properties of matter in the interior~\cite{Hinderer2008,AbbottPRL119,AbbottPRL121,AbbottPRX}. Recent detections of NS-NS (BNS) collisions (GW170817) or NS-BH mergers (GW200105 and GW200115) by the LIGO-Virgo-KAGRA collaboration of GW detectors have opened up new frontiers in multi-messenger astronomy ~\cite{AbbottAJL848}.
\\

In the context of GW, the secular quasi-normal modes (QNM) of NS are particularly interesting, since they carry information about the interior composition and viscous forces that damp these modes. QNMs in neutron stars are categorized by the restoring force that bring the perturbed star back to equilibrium ~\cite{Cowling,Schmidt,Thorne}. Examples include the fundamental $f$-mode, $p$-modes and $g$-modes (driven by pressure and buoyancy respectively), as well as $r$-modes (Coriolis force) and pure space-time $w$-modes. Several of these modes are expected to be excited during SN explosions~\cite{Radice_2019}, or in a starquake~\cite{Keer2014} or in isolated perturbed NSs ~\cite{Doneva}, or during the post-merger phase of a binary NS ~\cite{Stergioulas2011,Bauswein2012,Takami2014}, with the $f$-mode being the primary target of interest. It has been argued that spin and eccentricity enhance the excitation of
the $f$-modes during the inspiral phase of NS mergers ~\cite{Chirenti_2017,Steinhoff2021}. The fundamental $f$-modes are within the sensitivity range of current generation of GW detectors and are correlated with the tidal deformability during the inspiral phase of NS mergers ~\cite{Chan2014,Hinderer2016,Pratten2020,Andersson2021}.The g-modes  can be excited during inspiral of a merger event  ~\cite{Andersson2021} and are also sensitive to the internal composition of NS~\cite{Constantinou2021,Zhao2022a}. However, the impact on GW is too weak to  be noticed  by the current generation of instruments~\cite{Andersson2021}. Which leads us to focus on $f$-mode oscillation of NS.
\\

Among the many studies in the literature that study the $f$-mode, the pioneering work of Andersson and Kokkotas ~\cite{Andersson96,Andersson98} relating the NS global properties such as mass, radius or compactness with the frequency and damping times of QNMs is the most relevant motivation to our work. However, the majority of these studies rely on the Cowling approximation (neglecting perturbations of the background metric), instead of calculating in full general relativity (GR). While the Cowling approximation is justified as a first reasonable estimate of the mode frequency, full GR is required for a more accurate computation of the mode frequency and to find the damping time, in order to extract reliable information about the NS EoS from GW data.
\\

In a recent study~\cite{Sukrit}, we performed a systematic investigation of the role of nuclear saturation parameters on the oscillation modes for a purely nucleonic non-rotating NS in the framework of the RMF model. We then extended this investigation ~\cite{Pradhan2021} to study the effect of the appearance of hyperons on the $f$-mode frequencies. Completing the analysis, in this work, we present the results of calculations of $f$-modes of hyperonic stars in a fully general relativistic framework. 
\\

This paper is organized in the following way. In \cref{sec:formalism}, we discuss the RMF model Lagrangian and the model's parameters. In \cref{sec:macro}, the resulting macroscopic properties of the NS are presented, followed by \cref{sec:GR} detailing the GR equations used to determine the global $f$-mode frequency. We compile our results in \cref{sec:results} and summarize our conclusions in \cref{sec:discussions}.

\section{Microscopic model for the Equation of State}
\label{sec:formalism}

\subsection{The Relativistic Mean Field (RMF) Model}
The charge-neutral, beta equilibrated matter in the NS interior is described by our chosen RMF theory, which provides a Lorentz covariant description of the micro-physics of the NS interior. In the RMF model, baryon-baryon interaction is mediated by the exchange of scalar   ($\sigma$), vector   ($\omega$) and isovector ($\rho$) mesons, while hyperon-hyperon interactions are mediated by additional strange scalar   ($\sigma^*$), and strange vector ($\phi$) mesons ~\cite{Schaffner96}. The interaction Lagrangian density ($\mathcal{L}$) can be written as,

\begin{eqnarray}
     \mathcal{L} &=&\sum_B  \bar{\psi}_{_B}  \Big(i\gamma^{\mu}\partial_{\mu}-m_{_B}+g_{\sigma B}\sigma-g_{\omega B}\gamma_{\mu}\omega^{\mu}\nonumber \\
     &-&g_{\rho B}\gamma_{\mu} \vec{I_B}.\vec{\rho}^{\mu}\Big)\psi_{_B} 
     +\frac{1}{2}  (\partial_{\mu} \sigma \partial^{\mu}\sigma - m_{\sigma}^2 {\sigma}^2)-U_{\sigma}\nonumber \\
     &+&\frac{1}{2}m_{\omega}^2 \omega_{\mu}\omega^{\mu}-\frac{1}{4} \omega_{\mu \nu}\omega^{\mu \nu} -\frac{1}{4}  (\vec{\rho}_{ \mu \nu}.\vec{\rho}^{\mu \nu}-2 m_{\rho}^2 \vec{\rho}_{\mu}.\vec{\rho}^{\mu}) \nonumber \\
    &+&\Lambda_{\omega} (g_{\rho N}^2  \vec{\rho}_{\mu} .\vec{\rho}^{\mu}) \  (g_{\omega N}^2 \omega_{\mu}\omega^{\mu})  \nonumber \\ 
     &+&\mathcal{L}_{YY}+\mathcal{L}_{\ell} \,,
     \label{eqn:lagr}
\end{eqnarray}
where
 \begin{eqnarray}
     U_{\sigma}&=&\frac{1}{3}b m_N   (g_{\sigma N} \sigma)^3+\frac{1}{4}c   (g_{\sigma N} \sigma)^4\,, \nonumber\\
     \mathcal{L}_{YY}&=&\sum_Y  \bar{\psi}_{_Y}  (g_{\sigma^* Y} \sigma^*-g_{\phi Y}\gamma_{\mu}\phi^{\mu})\psi_{_Y}+\frac{1}{2}m_{\phi}^2 \phi_{\mu}\phi^{\mu}\nonumber\\
     && \ \ -\frac{1}{4} \phi_{\mu \nu}\phi^{\mu \nu}  +\frac{1}{2}  (\partial_{\mu} \sigma^* \partial^{\mu}\sigma^* - m_{\sigma^*}^2 {\sigma^*}^2)\,, \nonumber\\
    \mathcal{L}_{\ell}&=&\sum_{\ell=\{e^- , \ \mu^-\}}  \bar{\psi}_{\ell}  (i\gamma^{\mu}\partial_{\mu}-m_{\ell}){\psi}_{\ell}\,. \nonumber
\end{eqnarray}
The governing field equations for constituent baryons and mesons can be found in our previous work~\cite{Pradhan2021}. In the mean-field approximation, meson fields are replaced by their ground state expectation values. Replacing the non-vanishing mean-meson expectation components as~\cite{chen2014}, `$\bar{\sigma}=\langle\sigma\rangle,\ \bar{\omega}_0= \langle \omega_0 \rangle, \ \bar{\rho}_{03}=\langle \rho_{03} \rangle, \ \bar{\sigma^*}=\langle \sigma^* \rangle,\ \bar{\phi}_0=\langle \phi_0 \rangle$', the energy density for the given Lagrangian density ~\eqref{eqn:lagr} is given by~\cite{Pradhan2021}:
\begin{eqnarray}
    \epsilon &=&\frac{1}{2}m_{\sigma}^2\bar{\sigma}^2+\frac{1}{2}m_{\sigma^*}^2\bar{\sigma^*}^2+\frac{1}{2}m_{\omega}^2\bar{\omega}_0^2+\frac{1}{2}m_{\rho}^2\bar{\rho}_{03}^2\nonumber\\
    &+& \frac{1}{2}m_{\phi}^2\bar{\phi}_0^2+\frac{1}{3}b m_N   (g_{\sigma N} \bar{\sigma})^3+\frac{1}{4}c   (g_{\sigma N} \bar{\sigma})^4 \nonumber \\
    &+&\sum_B \frac{g_{sB}}{2\pi^2} \  
    \int_0^{k_{FB}} {\sqrt{k^2+{m_B^*}^2}}\  dk \nonumber\\
    &+& 3 \Lambda_{\omega} (g_{\rho N} g_{\omega N} \bar{\rho}_{03} \bar{\omega}_0 )^2 \nonumber\\
    &+& \sum_{\ell} \frac{g_{s\ell}}{2\pi^2} \ \int_0^{k_{F{\ell}}} {\sqrt{k^2+{m_{\ell}}^2}}\  dk \,, \label{eqn:endens}
\end{eqnarray}
 where $g_{si}$ and $k_{Fi}$ represent spin degeneracy and Fermi momentum of  $i^{th}$ species respectively. $m_B^*$ is the effective  mass for baryon $B$ and given by,
\begin{eqnarray}
    m^*_B &=& m_B-g_{\sigma B}  \bar{\sigma} - g_{\sigma^*B}  \bar{\sigma}^*\,. 
    \label{eqn:effmass}
\end{eqnarray}
 
The pressure ($p$) is given by the Gibbs-Duhem relation~\cite{Hornick}
\\
 \begin{equation}
     p=\sum_{i=B,\ell} \mu_i n_i-\epsilon 
     \label{eqn:pres}
 \end{equation}
 with $n_i$ and $\mu_i$ as the number density and chemical potential of the $i^{th}$ constituent respectively. The baryon and lepton chemical potentials can be expressed respectively as,
 \begin{eqnarray}
   \mu_B &=& \sqrt{k_{FB}^ 2  + {m^*_B}^2 }  + g_{\omega_B} \bar{\omega}_0 +g_{\phi_B} \bar{\phi}_0+I_{3_B} g_{\rho_B} \bar{\rho}_{03}\,~, \nonumber \\
   \mu_{\ell}&=&\sqrt{k_{F\ell}^ 2  + {m_{\ell}}^2}\,.
    \label{eqn:chempot}
\end{eqnarray}

\subsection{Parameters of the RMF model}
\label{subsec:para}

\subsubsection{Nucleon Couplings}
Here, we briefly discuss the coupling constants, which may be regarded as model parameters. The nucleon isoscalar coupling constants  ($g_{\sigma N},g_{\omega N},b,c$) are set by fixing nuclear saturation properties: nuclear saturation  density ($n_0$), binding energy per nucleon  ($E/A$ or $E_{\rm sat}$), incompressibility  ($K$) and the effective nucleon mass ($m^*$) at saturation. The isovector coupling constants  ($g_{\rho N},\Lambda_{\omega}$) are obtained by fixing the symmetry energy  ($J$), and its slope  ($L$) at saturation~\cite{Hornick,chen2014}. It was concluded that in RMF models the stiffness of the EoS is mainly controlled by $m^*$~\cite{Hornick}. We consider a reasonable range of $m^*$ such that the maximum mass is above the observed limit ($m^*<0.75$) and does not induce the appearance of instabilities in the neutron matter EoS ($m^*>0.55$)~\cite{Hornick}. The effect of astrophysical constraints on $m^*$ is discussed in detail at the end of~\Cref{sec:macro}~. The range of saturation nuclear parameters considered in this work, have been summarized in ~\Cref{tab:rangepara}. 
\begin{table}[h]
    \centering
\begin{tabular}
{|p{1.1 cm}|p{1.cm}|p{1.1cm}|p{1.1cm}|p{1cm}|p{1cm}|p{1.1cm}|}
\hline
    Model & $n_0$ & $E_{sat}$& $K$& $J$ & $L$& $m^*/m_N$\\
      &($fm^{-3}$) & (MeV) & (MeV) & (MeV) & (MeV)& \\
\hline
 \hline
    
    RMF~\cite{Hornick}&0.15-0.16&  -15.5  $\pm$0.5& 230-280 &  32$\pm$2 & 50$\pm$10 & 0.65  $\pm$0.10\\
    
    \hline
\end{tabular}
\caption{Range of saturation nuclear parameters used in this work. Masses of mesons and the nucleon are fixed as $m_{\sigma}=550 \ \rm{ MeV}, \ m_{\omega}=783\  \rm{ MeV}, \ m_{\rho}=770\  \rm{MeV}, \ m_{\sigma^*}=975\  \rm{MeV}, \ m_{\phi}=1020\  \rm{MeV} \text{ and } \ m_N=939\  \rm{MeV} $. For the masses of the hyperons, we use data from~\cite{PDG2020}. }   
\label{tab:rangepara}
\end{table}
\subsubsection{Hyperon Couplings}
RMF models with attractive hyperon-hyperon interaction (mediated  by the strange meson $\sigma^*$) are incompatible with the current highest observed NS mass~\cite{ChatterjeeNPA}. Thus, we exclude the attractive hyperon-hyperon interaction. The non-strange scalar-hyperon couplings ($g_{\sigma Y}$) are fitted to available hyperon-nucleon potential depth in normal nuclear matter ($U_Y^N (n_0)$) using Eq.~\eqref{eqn:hyppot} ~\cite{Schaffner96,ChatterjeeNPA} and the vector and iso-vector hyperon couplings ($g_{\omega Y},g_{\rho Y},g_{\phi Y}$) are fixed to their theoretical values using the symmetries of the SU(6) quark model summarised in~\cite{Schaffner96,ChatterjeeNPA}.
\begin{equation}
    U_{Y}^{N} (n_0)=-g_{\sigma Y}\bar{\sigma}+g_{\omega Y}\bar{\omega_0}\,.
    \label{eqn:hyppot}
\end{equation}
\begin{eqnarray}
    g_{\omega {\Lambda}}=g_{\omega {\Sigma}}=2g_{\omega {\Xi}}&=&\frac{2}{3}g_{\omega N} \,,\nonumber\\
g_{\rho {N}} = g_{\rho {\Xi}} = \frac{1}{2} g_{\rho {\Sigma}} \> ; \> & g_{\rho {\Lambda}}&=0 \,,\nonumber\\
    2g_{\phi {\Lambda}}= 2g_{\phi {\Sigma}}=g_{\phi {\Xi}}&=&\frac{-2\sqrt{2}}{3}g_{\omega N}\,.
    \label{eqn:su6}
\end{eqnarray}

Among the nucleon-hyperon potentials $U_Y^N$, the best known potential depth is that of $\Lambda$, $U_{\Lambda}^N(n_0)=-30 \ \rm{MeV}$ ~\cite{Millener88,Schaffner92}. Although there is an uncertainty in $U_{\Sigma}^N$, it has been concluded from  experiments that $U_{\Sigma}^N$ is repulsive ~\cite{Schaffner92,Mares95,SchaffnerGal00,FriedmanGal07}. We fix the $U_{\Sigma}^N$ potential to its most commonly adopted value +30 MeV. However, the value of $U_{\Xi}^N$ is attractive and highly uncertain~\cite{ Fukuda98,Khaustov00,SchaffnerGal00}. So we vary the value of $U_{\Xi}^N$ within the range of -40 MeV to +40 MeV for our investigation. Once all the coupling constants for a fixed parameters set are determined, the EoS can be evaluated for the Lagrangian given in Eq.\eqref{eqn:lagr}.
 Each of the saturation parameters is  randomly drawn  from a uniform distribution defined in the range of  the corresponding parameter as given in \Cref{tab:rangepara}. After applying the astrophysical constraints ( $M_{\rm max}\geq 2M_{\odot}$ and tidal deformability constraint from GW170817~\cite{AbbottPRX} ), We are left with approximately 1500 (1483 to be exact) microscopic models for pure nucleonic matter and 1000 (1123 to be exact) microscopic models for neutron stars with nucleon-hyperon matter. For providing the asterosismolgy relations in \cref{sec:results}, we have obtained $f$-mode characteristics for ~$2.5\times 10^{5}$ neutron star.
\section{Macroscopic features of the Neutron Star}
\label{sec:macro}
After the EoS is specified,
the macroscopic structure of the NS can be  described by solving the Tolman–Oppenheimer–Volkoff (TOV) equations. Starting with general spherically symmetric metric \eqref{eqn:metric}, the equations describing  hydrostatic equilibrium (TOV)  are given by Eqs.~\eqref{tov1}-\eqref{tov2} and the equations governing metric 
functions $\Phi(r)$ and $\lambda (r)$ by Eqs.~\eqref{phi}-\eqref{lambda}.
\begin{equation}
     ds^2=-e^{2\Phi (r)}dt^2+e^{2\lambda (r)}dr^2+r^2 d\theta^2+r^2\sin^2{\theta} d\phi^2 \label{eqn:metric}
 \end{equation}
\begin{eqnarray} 
    \frac{dm(r)}{dr}&=& 4\pi r^2 \epsilon (r)  \label{tov1}\\
    \frac{dp(r)}{dr}&=&- \left[p (r)+\epsilon (r)\right]\frac{m (r)+4\pi r^3p (r)}{r (r-2m (r))} \label{tov2} \\
     \frac{d \Phi(r)}{dr}&=&\frac{-1}{\epsilon(r)+p(r)}\frac{dp}{dr} \label{phi}\\
     e^{2\lambda (r)}&=& \frac{r}{r-2m(r)} \,.\label{lambda}
\end{eqnarray}
Integration of TOV equations for a given EoS ($p=p(\epsilon)$) from the centre to the surface with vanishing pressure at the surface $p(R)=0$ provides the stellar radius $R$ and mass $M=m(R)$ for equilibrium stellar models. Another boundary condition is that at the surface, $\Phi(R)=\frac{1}{2}\log \l({1-\frac{2M}{R}}\r)$. The tidal love number $k_2$ for a given EoS can be evaluated by solving a set of additional differential equations along with TOV equations~\cite{YagiYunes2013}, which then lead to determination of another important observable quantity, the dimensionless tidal deformability (${\bar{\Lambda}}$) 
\begin{equation}
     \bar{\Lambda}=\frac{2}{3}k_2 \l(\frac{R}{M}\r)^5
    \label{eq:tidal_deformability}~.
\end{equation}

\begin{figure}[htbp]
    \centering
    \includegraphics[width=\linewidth]{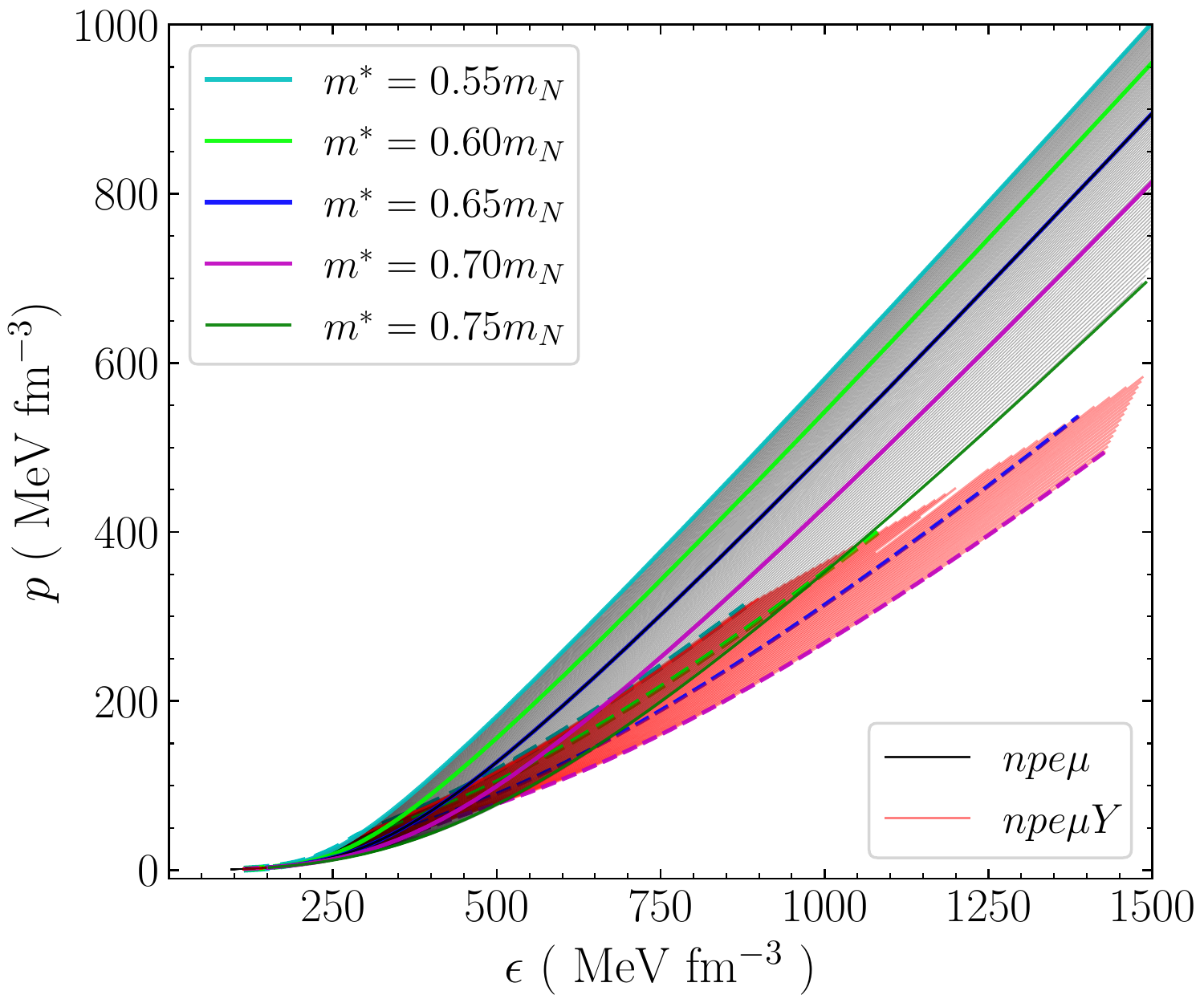}
    \caption{EoSs used in this work. To avoid cluttering we have displayed here a few of the EoSs with nucleons (solid black lines, $npe\mu$) as well as models with hyperons (solid red lines, $npe\mu Y$) corresponding to the parameter space explained in \cref{subsec:para}. Specific solid (dashed) lines are shown  for different effective mass for nucleonic (hyperonic) EoSs with other parameters fixed at $n_0=0.150\ \rm{fm^{-3}},\ E_{sat}=-16.0\ \rm{MeV}, \ J=32\ \rm{MeV}, \ L=60\ \rm{MeV}, \ K=240  \ \rm{MeV} \text{ and } U_{\Xi}=-18~\rm{MeV}$.}
    \label{fig:alleos}
\end{figure}

\begin{figure}[htbp]
    \centering
    \includegraphics[width=\linewidth]{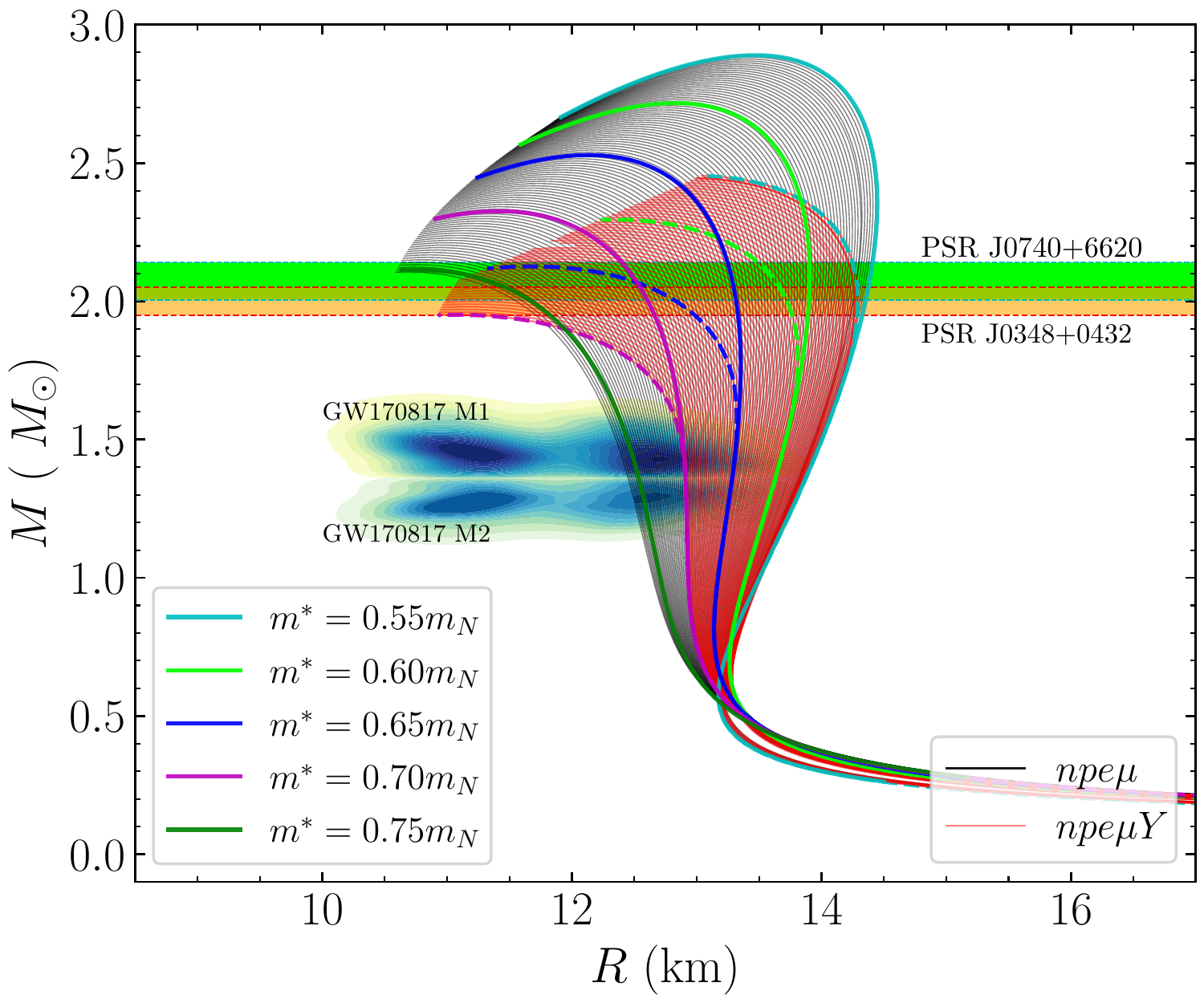}
    \caption{ Mass-Radius relation corresponding to EoSs used in this work (see \Cref{fig:alleos}). Horizontal bands correspond to masses $M=2.072^{+0.067}_{-0.066} M_{\odot}$ of PSR J0740+6620 ~\cite{riley2021} (green coloured band) and $M=2.01^{+0.04}_{-0.04} M_{\odot}$ of PSR J0348$+$0432 ~\cite{Antoniadis2013} (yellow coloured band). The mass radius estimates of the two companion neutron stars in the merger event GW170817~\cite{AbbottPRL121} are shown by shaded area labeled with GW170817 M1 (M2) \footnote{\url{https://dcc.ligo.org/LIGO-P1800115/public}}. }
    \label{fig:mr_withgw}
\end{figure}

We display the EoSs and corresponding mass-radius relations used in this work in \Cref{fig:alleos} and \Cref{fig:mr_withgw} respectively. Out of the EoSs obtained by randomly varying the saturation parameters in Table~\ref{tab:rangepara}, we only consider those which are compatible with recent astrophysical observational constraints, i.e., the EoS must reproduce the maximum observed neutron star of $2M_{\odot}$ mass and also be compatible with the tidal deformability estimation from the merger event GW170817~\cite{AbbottPRX}. The dominant parameter controlling the stiffness of the EoS is found to be the nuclear effective mass $m^*$~\cite{Sukrit}, and its range may be constrained by the maximum observed mass and compactness~\cite{David2020}. We note here that while considering models with nucleonic matter, the maximum $2M_{\odot}$ limit does not put any constraint on the uncertainty of $m^*$ while imposing the constraint of an upper limit of tidal deformability coming from GW170817~\cite{AbbottPRX} allows us to put a tight constraint on the lower limit, $m^* > 0.60m_N$. In the case of models with hyperonic EoSs, the maximum $2M_{\odot}$ puts a constraint on the upper limit $m^*\leq 0.70m_N$ along with the lower limit constraint  $m^*>0.60m_N$ from the upper limit of tidal deformability coming from GW170817. We found that EoS models satisfying the tidal deformability constraint from event GW170817 also satisfy NS radius constraint resulting from recent NICER observations~\cite{Miller2019,Riley2019}.


\section{Calculation of oscillation modes}
\label{sec:GR}
The theory of perturbed NSs, emitting  GWs at the characteristic frequency of its quasi-normal modes (QNM), was introduced in a paper by Thorne and Campolattaro in 1967~\cite{Thorne}. Many works~\cite{Sandoval,Flores2017}, including our previous work~\cite{Pradhan2021}, use the simplification defined by the relativistic Cowling approximation, where background metric perturbations are neglected. Frequencies obtained using the Cowling approximation for fundamental modes ($f$-modes) are purely real and differ by 20\%-30\% compared to frequencies obtained from the linearized equations of general relativity~\cite{Chirenti2015}. The Cowling approximation precludes a calculation of the damping time of QNMs. To obtain solutions in the fully general relativistic framework, different methods such as resonance matching (developed by  Thorne~\cite{ThorneIII} and later by Chandrasekhar~\cite{Chandrasekhar:1991}), direct integration~\cite{Detweiler83,Detweiler85}, and the method of continued fractions~\cite{Leins1993,Sotani2001}  have been applied.\\

Complicating effects like rotation are essential for describing a realistic astrophysical scenario. Recent efforts suggest that the leading order spin correction to the mode frequency is ~0.2($\nu_s/\nu_K$) ~\cite{Kruger2021} ($\nu_s$ is the spin frequency, and $\nu_K$ is the Kepler frequency). Almost all glitching pulsars have a low spin frequency ($\nu_s <$100 Hz). In contrast, the Kepler frequency is $\sim$1 kHz, such that they would have $f$-mode frequency correction $<2\%$, this implies that the rotation has a minor effect on a detection event from transient NS $f$-modes from glitching pulsars. For a merger scenario, recent efforts are going on to include the impact of $f$-mode dynamical tides and the spin effect on NS $f$-mode dynamical tide~\cite{Steinhoff2021}. A recent article ~\cite{Kuan2022} concludes that the spin correction to $f$-modes has a considerable impact on the gravitation wave for rapidly rotating stars. However, the effect of rotation is still a matter of investigation. \\

In this article, we employ the procedure developed by Lindblom and Detweiler (hereafter called LD) ~\cite{Detweiler83, Detweiler85} for finding the QNMs of the $f$-modes for non-rotating NSs. In short, the perturbation equations are solved inside the star with appropriate boundary conditions. Then a search for the complex QNM frequency ($\omega$)  is carried out for which one has only outgoing GWs at infinity. The real part (Re($\omega$)) of the obtained complex QNM frequency relates to the QNM frequency($f$) as ${\rm{Re}}(\omega)=2\pi f$, and the imaginary part represents the reciprocal of the damping time ($\frac{1}{\tau_f}$), i.e., the obtained  QNM frequency ($\omega$) has the form, $\omega=2\pi f+i\frac{1}{\tau_f}$. In this section, we present the basic equations that need to be solved for finding the complex QNM frequencies.
\subsubsection{Perturbations Inside the Star}
The perturbed metric ($ds^2_p$) can be written as ~\cite{Thorne},
\begin{eqnarray}
    ds^2_p=ds^2+h_{\mu \nu} dx^{\mu}dx^{\nu}~.
    \label{eqn:perturbedmetric}
\end{eqnarray}

Following the arguments given in Thorne and Campolattaro~\cite{Thorne}, we focus on the even-parity (polar) perturbations for which the the GW and matter perturbations are coupled. Then $h_{\mu \nu}$ can be expressed  as ~\cite{Sotani2001,Thorne},
\begin{eqnarray}
    h_{\mu \nu}=
   \begin{pmatrix}
r^lHe^{2\Phi} & i\omega r^{l+1} H_1 &0&0\\
i\omega r^{l+1} H_1 & r^l H e^{2\lambda} &0 &0\\
0 &0 & r^{l+2}K& 0\\
0 & 0& 0 &  r^{l+2}K sin^2{\theta}
\end{pmatrix} Y^l_m e^{i\omega t} ~, \nonumber\\
 \text{ }
 \label{eqn:metricfunctions}
\end{eqnarray}
Where $Y_m^l$ are spherical harmonics. $H,\ H_1, \ K$ are  perturbed metric functions and vary with $r$ (i.e.,  $H=H(r),\ H_1=H_1(r), \ K=K(r)$ ). The  Lagrangian displacement vector  $\textbf{$\zeta$}=(\zeta^r,\zeta^{\theta},\zeta^{\phi})$  associated with the polar perturbations of the fluid can be characterised as~\cite{Detweiler85,Tonetto2021},

\begin{eqnarray}
    \zeta^{r}&=&\frac{r^l}{r}e^{- \lambda} W(r)  Y^l_m e^{i\omega t} \nonumber \\
    \zeta^{\theta}&=&\frac{-r^l}{r^2} V(r)  \frac{\partial Y^l_m}{\partial \theta} e^{i\omega t} \nonumber \\
    \zeta^{\phi}&=&\frac{-r^l}{r^2 sin^2\theta} V(r)  \frac{\partial Y^l_m}{\partial \phi} e^{i\omega t}
    \label{eqn:pertfluid}
\end{eqnarray}

where $W,V$ are amplitudes of the radial and transverse fluid perturbations. The equations governing these perturbation functions and the metric perturbations inside the star are given by~\cite{Sotani2001,Tonetto2021},
\begin{eqnarray}
    \frac{d H_1}{dr}&=&\frac{-1}{r}\left[l+1+\frac{2m}{r}e^{2\lambda}+4\pi r^2e^{2\lambda} \left( p-\epsilon \right)\right] H_1 \nonumber\\
    &+&\frac{1}{r}e^{2\lambda}\left[H+K+16\pi\left(p+\epsilon\right)V\right] \label{eqn:dh1} \ , \\
    \frac{d K}{dr}&=&\frac{l\l(l+1\r)}{2r}H_1+\frac{1}{r}H-\l(\frac{l+1}{r}-\frac{d\Phi}{dr}\r)K \nonumber \\
    &+&\frac{8\pi}{r}\l(p+\epsilon\r)e^{\lambda} W \ ,  \label{eqn:dk} \\
    \frac{d W}{dr}&=&re^{\lambda}\l[\frac{1}{\gamma p}e^{-\Phi}X-\frac{l\l(l+1\r)}{r^2}V-\frac{1}{2}H-K\r] \nonumber \\
    &-&\frac{l+1}{r}W   \label{eqn:dw} \ , \\
    \frac{d X}{dr}&=& \frac{-l}{r}X+\l(p+\epsilon \r)e^{\Phi}\Bigg[\frac{1}{2}\l(\frac{d\Phi}{dr}-\frac{1}{r}\r)H \nonumber\\
    &-&\frac{1}{2}\l( \omega^2re^{-2\Phi}+\frac{l(l+1)}{2r}\r)H_1+\l(\frac{1}{2r}-\frac{3}{2}\frac{d\Phi}{dr}\r)K \nonumber\\
    &-&\frac{1}{r}\l[ \omega^2\frac{e^{\lambda}}{e^{2\Phi}}+4\pi \l(p+\epsilon \r) e^{\lambda}-r^2\frac{d}{dr}\l( \frac{e^{-\lambda}}{r^2}\frac{d \Phi}{dr}\r)\r]W \nonumber \\
    &-&\frac{l(l+1)}{r^2}\frac{d\Phi}{dr}V\Bigg]  \label{eqn:dx}\ ,
\end{eqnarray}
\begin{eqnarray}
   &&\l[1-\frac{3m}{r}-\frac{l(l+1)}{2}-4\pi r^2p\r]H-8\pi r^2 e^{-\Phi}X \nonumber\\
   &-& \l[ 1+ \omega^2r^2e^{-2\Phi} -\frac{l(l+1)}{2}-(r-3m-4\pi r^3p)\frac{d\Phi}{dr}\r]K \nonumber \\
   &+&r^2e^{-2\lambda}\l[\omega^2e^{-2\Phi}-\frac{l(l+1)}{2r}\frac{d\Phi}{dr}\r]H_1 =0  \label{eqn:h}\\
  && e^{2\Phi}\l[ e^{-\phi}X+\frac{e^{-\lambda}}{r}\frac{dp}{dr} W+\frac{(p+\epsilon)}{2}H\r] \nonumber \\
  &-&\omega^2 \l(p+\epsilon\r) V=0 ~, \label{eqn:v}
\end{eqnarray}

where $X$ is introduced as~\cite{Detweiler83,Sotani2001}
\begin{eqnarray}
    X&=&\omega^2\l(p+\epsilon \r) e^{-\Phi} V-\frac{We^{\Phi-\lambda}}{r}\frac{dp}{dr}-\frac{1}{2} \l(p+\epsilon \r) e^{\Phi}H\,, \nonumber \\
    \text{}  \label{eqn:x}
\end{eqnarray}
$m=m(r)$ is the enclosed mass of the star and $\gamma$ is the adiabatic index defined as
\begin{equation}
    \gamma=\frac{\l(p+\epsilon \r)}{p}\l(\frac{\partial p}{\partial \epsilon }\r)\bigg|_{ad} ~.
     \label{eqn:gamma}
\end{equation}

While solving the differential equations Eqs. \eqref{eqn:dh1}-\eqref{eqn:dx} along with the algebraic Eqs. \eqref{eqn:h}-\eqref{eqn:v}, we have to impose proper boundary conditions, i.e., the perturbation functions are finite throughout the interior of the star (particularly at the centre, i.e., at $r=0$) and the perturbed pressure ($\Delta p$) vanishes at the surface.  Function values at the centre of the star can be found using the Taylor series expansion method described in Appendix B of~\cite{Detweiler83} (see also  Appendix A of~\cite{Sotani2001}. It is to be noted that the first term in RHS of Eq.~(A15) in ~\cite{Sotani2001} misses a factor $\epsilon$). The vanishing perturbed pressure at the stellar surface is equivalent to the condition $X(R)=0$ (as, $\Delta p=-r^le^{-\Phi}X$). We followed the procedure described in LD~\cite{Detweiler83} to find the unique solution for a given value of $l$ and $\omega$ satisfying all the boundary conditions inside the star.
\subsubsection{Perturbations outside the star and complex eigenfrequencies}
The perturbations outside the star are described by the Zerilli  equation~\cite{Zerilli}.
\begin{equation}
    \frac{d^2Z}{dr_*^2}+\omega^2 Z=V_Z Z
    \label{eqn:zerilli}
\end{equation}
 where $r_*=r+2M \log \l({\frac{r}{2M}-1}\r)$ is the tortoise co-ordinate and $V_Z$ is defined as ~\cite{Zerilli},
 \begin{eqnarray}
     V_Z&=&\frac{2\l(r-2M\r)}{r^4 \l(nr+3M\r)^2}\Big[n^2(n+1)r^3 \nonumber \\
     &+&3n^2Mr^2+9nM^2r+9M^3\Big] ~,
 \end{eqnarray}
where $n=\frac{1}{2} (l+2)(l-1)$. Asymptotically the wave solution to \eqref{eqn:zerilli} can be expressed as \eqref{eqn:zerillisolution},
\begin{eqnarray}
    Z=A(\omega)Z_{in}&+&B(\omega) Z_{out}\,, \label{eqn:zerillisolution}\\
    Z_{out}=e^{-i\omega r^*} \sum_{j=0}^{j=\infty}\alpha_j r^{-j}&,& Z_{in}=e^{i\omega r^*} \sum_{j=0}^{j=\infty}\bar{\alpha}_j r^{-j} ~.\nonumber
    \end{eqnarray}
 Keeping  terms up to $j=2$ one finds,
 \begin{eqnarray}
     \alpha_1&=&-\frac{i}{\omega}(n+1)\alpha_0, \\
     \alpha_2&=&\frac{-1}{2\omega^2}\l[n(n+1)-i3M\omega\l(1+\frac{2}{n}\r)\r]\alpha_0
 \end{eqnarray}

For initial boundary values of Zerilli functions, we use the method described in ~\cite{Fackerell,Detweiler85,Sotani2001}. Setting $m=M$ and perturbed fluid variables to 0 (i.e., $W=V=0$) outside the star, connection between the metric functions \eqref{eqn:metricfunctions} with Zerilli function ($Z$ in Eq.\eqref{eqn:zerilli}) can be written as,
\begin{eqnarray}
    \begin{pmatrix}
    r^lK\\
    r^{l+1}H_1
    \end{pmatrix}
    &=&Q\begin{pmatrix}
    Z \\
    \frac{dZ}{dr_*}
    \end{pmatrix}
    \label{eqn:zerillconnection}
\end{eqnarray}
\begin{eqnarray*}
    Q&=&\begin{pmatrix}
    \frac{ n(n+1)r^2+3nMr+6M^2}{r^2(nr+3M)} & 1\\
    \frac{nr^2-3nMr-3M^2}{(r-2M)(nr+3M)} & \frac{r^2}{r-2M}
    \end{pmatrix} \\
\end{eqnarray*}

The initial boundary values of Zerilli functions are fixed using \eqref{eqn:zerillconnection}. Then, the Zerilli equation \eqref{eqn:zerilli} is integrated numerically to infinity and the complex coefficients $A(\omega) ,\ B(\omega)$ are obtained matching the analytic  expressions for $Z$ and $\frac{dZ}{dr_*}$ with the numerically obtained value of $Z$ and $\frac{dZ}{dr_*}$. The natural frequencies of an oscillating neutron star, which are not driven by incoming gravitational radiation, represent the quasi-normal mode frequencies. Mathematically we have to find the complex roots of  $A(\omega)=0$, representing the complex eigenfrequencies of  QNMs.
\\

We tested our numerical technique by reproducing the quadrupole $f$-mode frequencies (complex)  from ~\cite{Sotani2001} for polytropic stars ( i.e., Table V and Figure. 5 of ~\cite{Sotani2001} wherein the    method of continued fractions was used to find the complex eigenfrequencies ).
    
\section{Results}
\label{sec:results}


\subsection{Universal relations in NS asteroseismology}\label{sec:universal_relations}

NS asteroseismology (inverse asteroseismology), the technique of inferring the NS parameters (internal composition) from QNM characteristics, was first introduced by Andersson and Kokkotas ~\cite{Andersson96, Andersson98}. Theoretically, it was shown that the frequency of $f$-mode varies linearly with density whereas, the damping time varies inversely with stellar compactness when scaled by $M^3/R^4$. Therefore empirical fit relations can be defined as follows:
\begin{equation}
 f(\text{kHz}) = a_r + b_r \sqrt{ \frac{{M}}{{{R}^3}} } 
 \label{eq:freqfit}
\end{equation}
\begin{equation}
 \frac{R^4}{M^3\tau_f} = a_i + b_i  \frac{{M}}{{{R}}}~.
 \label{eq:dampingtimefit}
 \end{equation}
where the constants $a_r,b_r,a_i,b_i$ are extracted from the best fit to the data. The fits  were subsequently improved by other works by including few selected realistic EoSs or those with exotic matter (hyperons and quarks) ~\cite{Benhar, Salcedo}. Further, NS rotation was considered by Doneva~et~al.~\cite{Doneva}, where empirical relations for frequency in non rotating limits are also given. However many of the selected EoSs considered in previous works are now incompatible with the $\approx 2M_{\odot}$ maximum mass constraint or with the tidal deformability (radius) constraints from the event GW170817, and are hence ruled out.
\\

\begin{table}[htbp]
   \centering
\begin{tabular}{|p{4cm}|p{1.2cm}|p{1.5cm}|}
\hline
    Reference   & $a_r$ (kHz) & $b_r$ (kHz$\times$km)  \\
\hline
\small{Andersson \& Kokkotas ~\cite{Andersson98}} & 0.22 & 47.51\\
 \hline
   Benhar \& Ferrari ~\cite{Benhar} & 0.79 & 33\\
  \hline
 D.Doneva et al ~\cite{Doneva} &1.562& 25.32\\
 \hline
 Pradhan \& Chatterjee ~\cite{Pradhan2021} & 1.075 & 31.10\\
  \hline
  This Work   & 0.535  &36.20\\
  \hline
\end{tabular}
\caption{Asteroseismology relation coefficients for $f$-mode frequency from different works. The coefficients $a_r$ and $b_r$ are related to $f$  by Eq.~\eqref{eq:freqfit}. }
\label{tab:fitrelations_freq}
\end{table}
\begin{table}[htbp]
   \centering
\begin{tabular}{|p{4cm}|p{1.2cm}|p{1.5cm}|}
\hline
    Reference   & $a_i$  & $b_i$   \\
\hline
\small{Andersson \& Kokkotas ~\cite{Andersson98}} & 0.086 & -0.267\\
 \hline
   Benhar \& Ferrari ~\cite{Benhar} & 0.087 & -0.271\\
  \hline
  This Work & 0.080 &  -0.245 \\
  \hline
\end{tabular}
\caption{Asteroseismology relation coefficients for $f$-mode damping time from different works. The coefficients $a_i$ and $b_i$ are related to $\tau_f$ by Eq.~\eqref{eq:dampingtimefit}.}
\label{tab:fitrelations_dampingtime}
\end{table}

\begin{figure}[htbp]
    \centering
    \includegraphics[width=\linewidth]{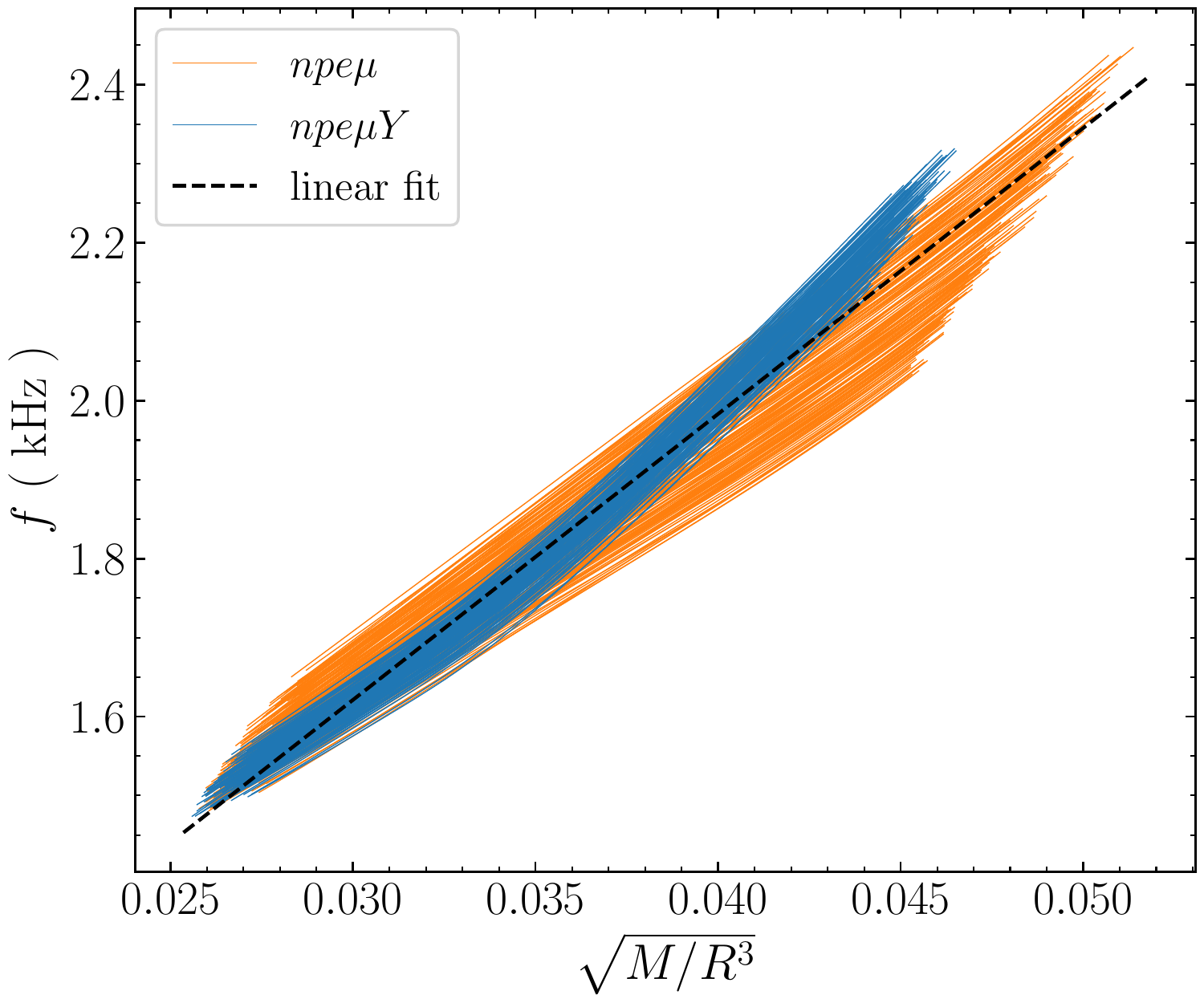}
    \caption{$f$-mode frequencies as a function of square root of the mean density. Models with only nuclear (nucleon and hyperon) matter are shown in orange (blue)  lines and the empirical linear fit relation \eqref{eq:freqfit} by black dashed line.}
    \label{fig:freqvsdensity}
\end{figure}

\begin{figure}[htbp]
    \centering
    \includegraphics[width=\linewidth]{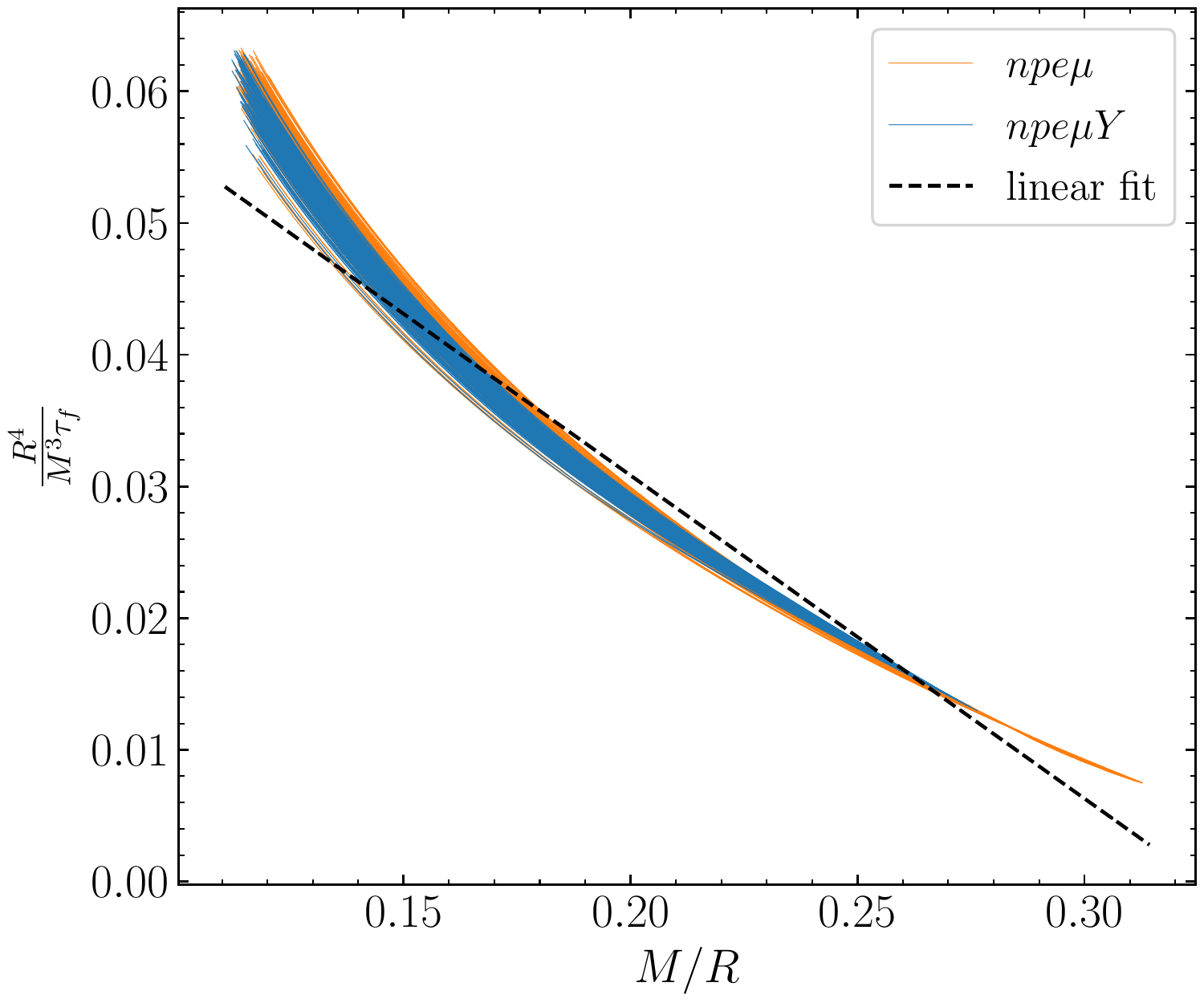}
    \caption{Scaled damping time of $f$-mode  as a function of stellar compactness $M/R$. Models with only nuclear (nucleon and hyperon) matter are shown in orange (blue) lines and the empirical linear fit relation \eqref{eq:dampingtimefit} by black dashed line.}
    \label{fig:dampingtimevscompactness}
\end{figure}

It is worth noting that although the empirical relations obtained previously aim to be independent of the underlying EoS, all the proposed empirical fit relations are somewhat model dependent. The knowledge of mode frequencies and the NS masses (which is among the most precisely determined global variables) can therefore help to discriminate among the different EoSs, or to understand the behaviour of high density NS matter ~\cite{Wen}. In other words, these empirical relations can be used not just to infer mass and radius, but also constrain the EoS stiffness and the presence of exotic matter ~\cite{Benhar}. Instead of choosing selected EoSs, we fit asteroseismology relations to cover the full range of uncertainties in nuclear and hypernuclear saturation parameters in the EoS subject to current astrophysical constraints. Empirical fit relations for frequency and damping time of $f$-modes from different works along with this work are tabulated in \Cref{tab:fitrelations_freq} and \Cref{tab:fitrelations_dampingtime} respectively. In this work we found, $a_r=0.535 \pm 7.383\times10^{-4},\ b_r=36.206 \pm 0.019, a_i= (7.99\pm 0.002)\times 10^{-2}$ and $b_i=-0.245 \pm1.005\times 10^{-4}$. 
 One may compare the fit results within Cowling approximation~\cite{Pradhan2021} and full GR calculations (this work) for $a_r$ and $b_r$ from Table~\ref{tab:fitrelations_freq}.
The dependence of frequency (scaled damping time) with density (compactness) is displayed in \Cref{fig:freqvsdensity} (\Cref{fig:dampingtimevscompactness}), along with empirical fit relation Eq.~\eqref{eq:freqfit} (Eq.~\eqref{eq:dampingtimefit}).
Please note that the coefficients given in \cite{Andersson98} were incorrect due to a normalisation error in the calculation \footnote{K. Kokkotas, private communication}. These values have now been updated in ~\Cref{tab:fitrelations_freq}. 
\\


\begin{figure}[htbp]
    \centering
    \includegraphics[width=\linewidth]{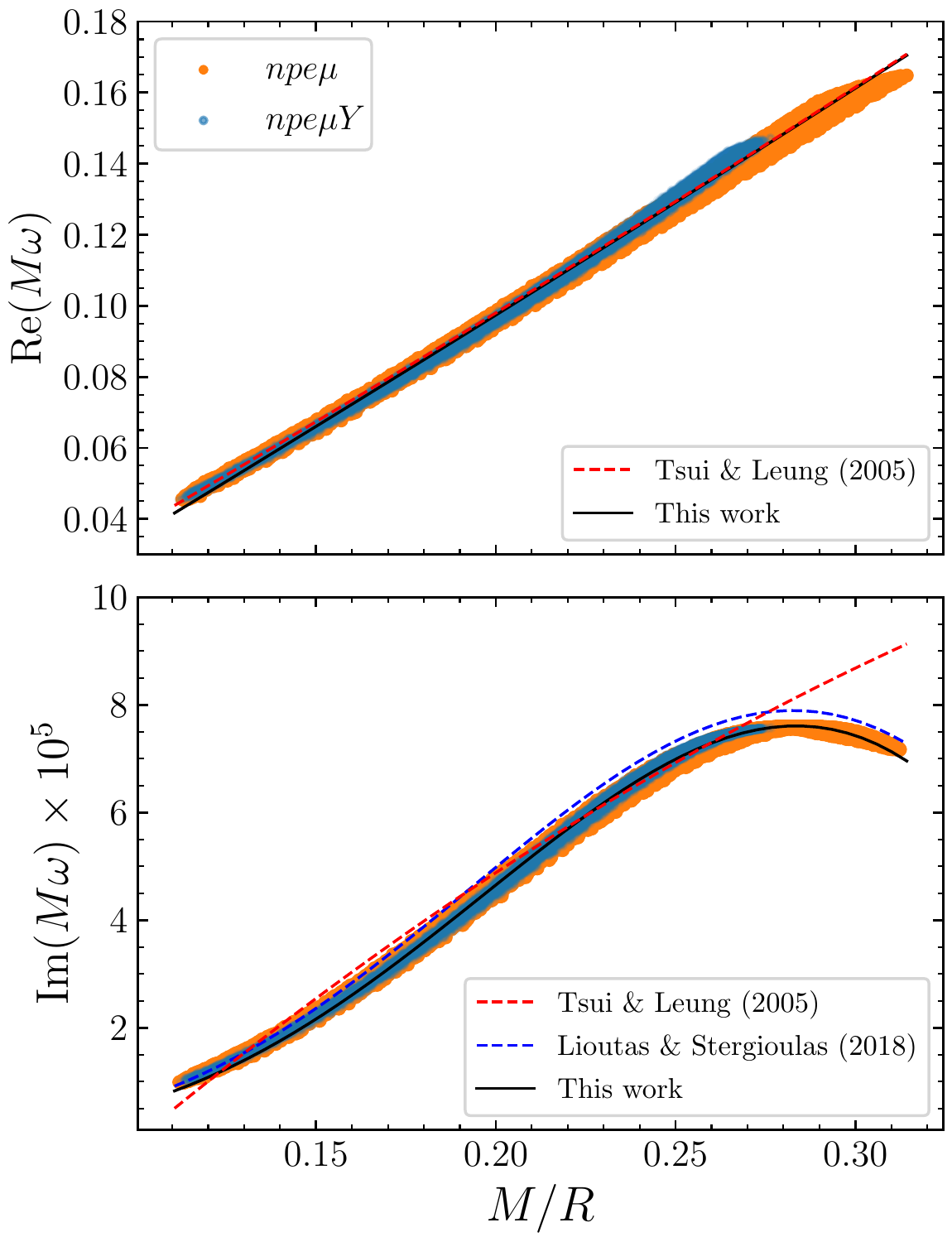}
    \caption{Dependence of QNM frequency after scaling with NS mass as a function of stelllar compactness. Scattered orange (blue) points correspond to models with $npe\mu $ ($npe\mu Y$) matter. The region spanned by hyperonic stars comes within the region spanned by the nucleonic models. Upper panel showing the universality for real part of $M\omega$, while lower panel shows the universality of mass scaled by damping time ($\frac{M}{\tau_f}$ or $\rm{Im}(M\omega)$). }
    \label{fig:momega_compactness}
\end{figure}

Contrary to empirical fit relations \eqref{eq:freqfit} and \eqref{eq:dampingtimefit} which are model dependent, there are other proposed universal relations (UR), which are fairly independent of underlying composition hence, more useful for extracting the NS parameter from QNM observables. It was shown when the mode characteristics are scaled with NS mass or radius they show correlations with the stellar compactness ~\cite{Andersson98} and the relations can be expressed in a universal way. In our previous work~\cite{Pradhan2021} we found the universality between scaled frequency with stellar compactness holds when $\omega$ scaled with NS mass but deviates from universality when scaled by radius. Tsui and Leung~\cite{Tsui2005} explicitly demonstrated that the scaled polar QNM frequencies of realistic neutron stars are approximately given by a universal function of the compactness, and improved the linear UR to quadratic fit, as given in Eq.~\eqref{eq:momega_universal_quadratic}. 
We note in \Cref{fig:momega_compactness} that a similar quadratic fit for Im($M\omega$) given in ~\cite{Tsui2005} deviates from universality at large compactness, whereas Eq.~\eqref{eq:universal_dampingtime} proposed by Lioutas and Stergioulas~\cite{Lioutas2017} provides a better fit.

 We display the dependence of scaled complex QNM frequency (scaled with NS mass), $M\omega$ as a function of compactness along with the URs from this and past works in \Cref{fig:momega_compactness}. Fit parameters corresponding to URs \eqref{eq:momega_universal_quadratic} and \eqref{eq:universal_dampingtime} found in this work are tabulated in \Cref{tab:momega_universalrelations}.
\\

\begin{equation}
    {\rm{Re}} (M\omega)=a_0\l(\frac{M}{R}\r)^2+a_1\frac{M}{R}+a_2
    \label{eq:momega_universal_quadratic}
\end{equation}
\begin{equation}
    {\rm{Im}}(M\omega)=b_0\l(\frac{M}{R}\r)^4+b_1\l(\frac{M}{R}\r)^5+b_2\l(\frac{M}{R}\r)^6
    \label{eq:universal_dampingtime}
\end{equation}
\begin{table}[htbp]

    \centering
    
    \begin{tabular}{|p{0.1\linewidth}|p{0.35\linewidth}|p{0.1\linewidth}|p{0.35\linewidth}|}
    \hline
 \multicolumn{2}{|c|}{$\rm{Re}(M\omega)$}& \multicolumn{2}{|c|}{$\rm{Im}(M\omega)$} \\
 \hline
         $a_0$&0.079$\pm 0.002$  &$b_0$& $\l(9.836\pm 0.003\r) \times 10^{-2} $ \\
         \hline
         $a_1$& 0.599 $\pm$ 0.001&$b_1$& $\l(-4.448\pm 0.002\r)\times 10^{-1}$\\
         \hline
         $a_2$&-0.026$\pm 8\times 10^{-5}$ &$b_2$&$\l(4.915\pm 0.004\r)\times 10^{-1}$\\
         \hline
    \end{tabular}
    \caption{Fit parameters for the URs \eqref{eq:momega_universal_quadratic} and \eqref{eq:universal_dampingtime} obtained in this work.}
    \label{tab:momega_universalrelations}
\end{table}

In a binary NS system, during the inspiral phase, NSs deform each other by exerting strong gravitational forces and the deformation depends upon the underlying EoS. The analysis of the tidal deformability from the event GW1701817 plays a crucial role in constraining NS EoS. From our current understanding of merger simulations, the mass scaled peak frequency ($f_{peak}$) of the post merger phase shows universality with tidal deformability or compactness ~\cite{Read2013,Bernuzzi2015,Vretinaris2020,Blacker2020}. It was pointed out recently by Chakravarti and Andersson~\cite{Kabir2020} that the universality between $f_{peak}$ and tidal deformability can be explained by adding the rotational and thermal corrections to the existing universal relation between mode frequency and tidal deformability of cold non-rotating neutron  stars  and the total  mass scaled $f_{peak}$ frequency can be expressed as a scaling factor times the mass scaled $f$-mode frequency.
Also during the inspiral phase the $f$-modes are most likely to be excited and observation of $f$-mode frequency along with tidal deformability can be used to probe the NS interior. Analyzing the event GW170817 along with the universality behaviour of frequency and tidal deformability should allow one to put a lower bound on the $f$-mode frequencies for NSs within the mass range of the two binary components of GW170817~\cite{Pratten2020}. 

\begin{figure}[htbp]
    \centering
    \includegraphics[width=\linewidth]{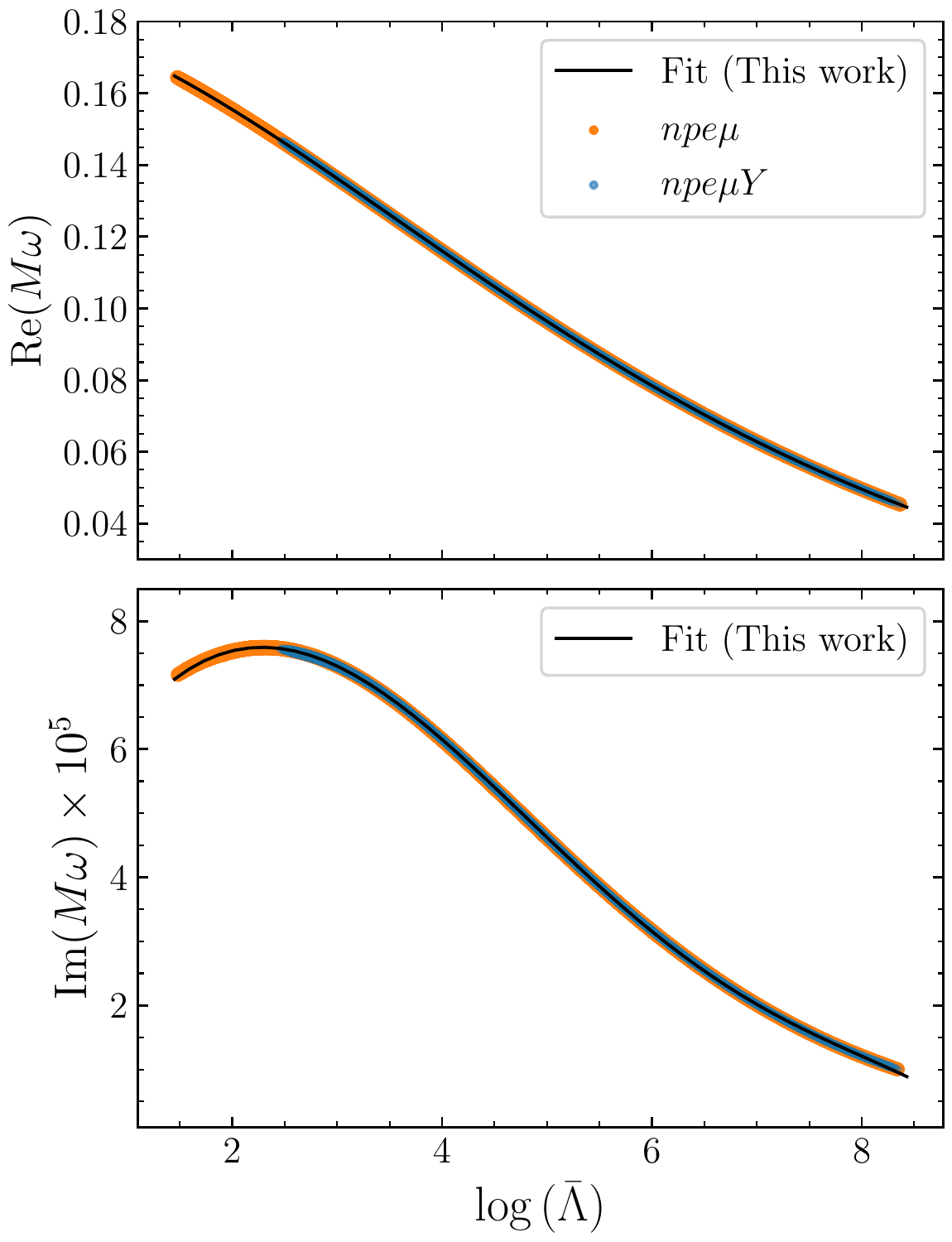}
    \caption{Dependence of QNM frequency after scaling with NS mass as a function of dimensionless tidal deformability. Scattered orange (blue) points correspond to model with $npe\mu$ ($npe\mu Y$) matter.The region spanned by hyperonic stars comes within the region spanned by the nucleonic models.  Upper panel showing the universality for real part of $M\omega$, while lower panel shows the universality of mass scaled by damping time ($\frac{M}{\tau_f}$ or $\rm{Im}(M\omega)$).}
    \label{fig:f_loverelation}
\end{figure}

It was argued that the $f$-mode frequencies can be detected very accurately with improved sensitivity of GW detectors, whereas damping time may not be detected with such good accuracy ~\cite{Kokkotas2001}; in this case the universal relations can be helpful in constraining the damping time. The detection of $f$-mode characteristics~\cite{Wen} or $f_{peak}$ ~\cite{Blacker2020} along with tidal deformability can also be used to verify the presence of quarks  in the interior of NS.  We provide the UR  between $f$-mode characteristics and tidal deformability as suggested in~\cite{Chan2014,Sotani_Bharat2021}. We tabulate the complex $\alpha_j$ from \eqref{eq:momega_tidal_universal} found in this work in \Cref{tab:f_love_fitparameters} and display in~\Cref{fig:f_loverelation}. 
\begin{equation}
    M\omega=\sum_j \alpha_j \l(\log{\bar{\Lambda}}\r)^j~.
    \label{eq:momega_tidal_universal}
\end{equation}

\begin{table*}[htbp]
    \centering
    \begin{tabular}{|p{3cm}|p{3cm}|p{3cm}|p{3cm}|p{3cm}|p{3cm}|}
    \hline
      $\alpha_0$   & $\alpha_1$ & $\alpha_2$ &$\alpha_3$ &$\alpha_4$ &$\alpha_5$  \\
   \hline
  $ 1.814\times 10^{-1}+ i3.362\times 10^{-5}$   & $-5.824\times 10^{-3}+ i3.993\times 10^{-5}$ & $-4.725\times 10^{-3}-i1.0215 \times 10^{-5}$ & $6.337\times 10^{-4}+i1.270\times 10^{-7}$ & $-2.871\times 10^{-5}+i1.230\times 10^{-7}$ & $3.150\times 10^{-7}-i7.817\times 10^{-9}$\\
   \hline
    \end{tabular}
    \caption{Values of the fit parameters $\alpha_j$  found in this work  for the given equation~\eqref{eq:momega_tidal_universal}.}
    \label{tab:f_love_fitparameters}
\end{table*}

We also found that there exists a universal relation between QNM characteristics (i.e., frequency and damping time) when they are scaled by NS mass. The universal relation between mass scaled angular frequency ($\rm{Re}(M\omega)$) and mass scaled damping time ($\rm{Im}(M\omega)$ or $M/\tau_f$) can be described by the following relation
\begin{equation}
    \rm{Im}(M\omega)=\sum_j \gamma_j \l(\rm{Re}(M\omega)\r)^j~.
    \label{eq:remomega_immomega_universal}
\end{equation}
We  tabulate the fit parameters of~\Cref{eq:remomega_immomega_universal} in~\Cref{tab:remomega_immomega_fitparameters}.\\

\begin{figure}
    \centering
    \includegraphics[width=\linewidth]{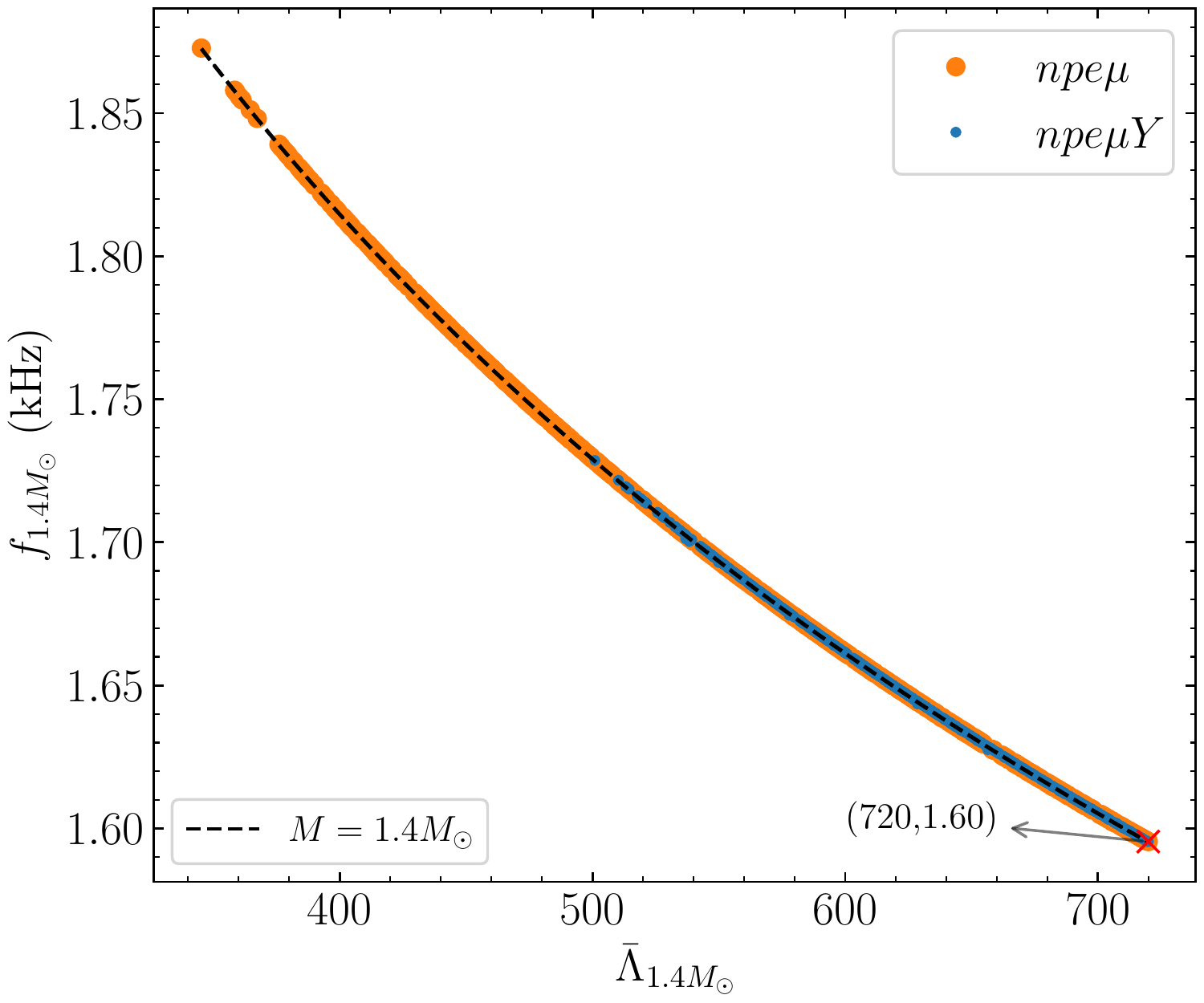}
    \caption{Universality between $f$-mode frequencies and tidal deformability ($\bar{\Lambda}$) of a canonical 1.4$M_{\odot}$ NS. Scattered orange (blue) points correspond to models with $npe\mu$ ($npe\mu Y$) matter. The black dashed line is obtained using the UR~\eqref{eq:momega_tidal_universal}. The red crossed point is corresponding to the maximum limit for $(\bar{\Lambda})_{1.4M_{\odot}}$.}
    \label{fig:f_tidal_1.4m}
\end{figure}

Even in a binary NS system, there exists universality between mode characteristics and tidal deformability in the inspiral phase~\cite{Wen,Chan2014}, and between $f_{peak}$ and tidal deformability~\cite{Blacker2020}. With future detections of BNS merger events the constraint on the tidal deformability will improve, and these in turn can then be used to constrain the mode characteristics. Our result for the lower bound on the mode frequency is in good agreement with the limit obtained from observations of GW170817~\cite{Pratten2020}. In order to test this hypothesis, we display the dependence of $f$-mode frequency  as a function of tidal deformability for canonical NSs in~\Cref{fig:f_tidal_1.4m}. Looking at the points corresponding to the maximum limit of the tidal deformability of a $1.4M_{\odot}$ in~\Cref{fig:f_tidal_1.4m}, one can conclude that for a $1.4M_{\odot}$ the lower bound  on mode frequency will be around 1.60kHz. Similarly we found the upper bound on $\tau_f$ for a $1.4M_{\odot}$ NS to be 0.28 sec. 

\begin{table*}[htbp]
    \centering
    \begin{tabular}{|p{3cm}|p{3cm}|p{3cm}|p{3cm}|p{3cm}|p{3cm}|}
    \hline
      $\gamma_0$   & $\gamma_1$ & $\gamma_2$ &$\gamma_3$ &$\gamma_4$ &$\gamma_5$  \\
   \hline
  $6.002	\times 10^{-6}$ & $-4.053	\times 10^{-4}$ & $1.339	\times 10^{-2} $& $-6.577	\times 10^{-2}$ &  $ 2.620	\times 10^{-1}$ & -1.072\\
   \hline
    \end{tabular}
    \caption{Values of the fit parameters $\gamma_j$  found in this work  for the given equation~\eqref{eq:remomega_immomega_universal}}
    \label{tab:remomega_immomega_fitparameters}
\end{table*}

\subsection{Correlation studies}
The uncertainty in the EoS and hence in mass-radius relations corresponding to uncertainty associated with nuclear and hypernuclear saturation data are discussed in \Cref{subsec:para}. Having tested our numerical scheme for complex $f$-mode frequencies, and obtained scaling relations with neutron star global parameters, we now extend our investigation to study the effect of microscopic (saturation) parameters on the $f$-mode observables ( $f$ and $\tau_f$ ). 
\\
\subsubsection{Nucleonic Matter}
\begin{figure}[htbp]
    \centering
    \includegraphics[width=\linewidth]{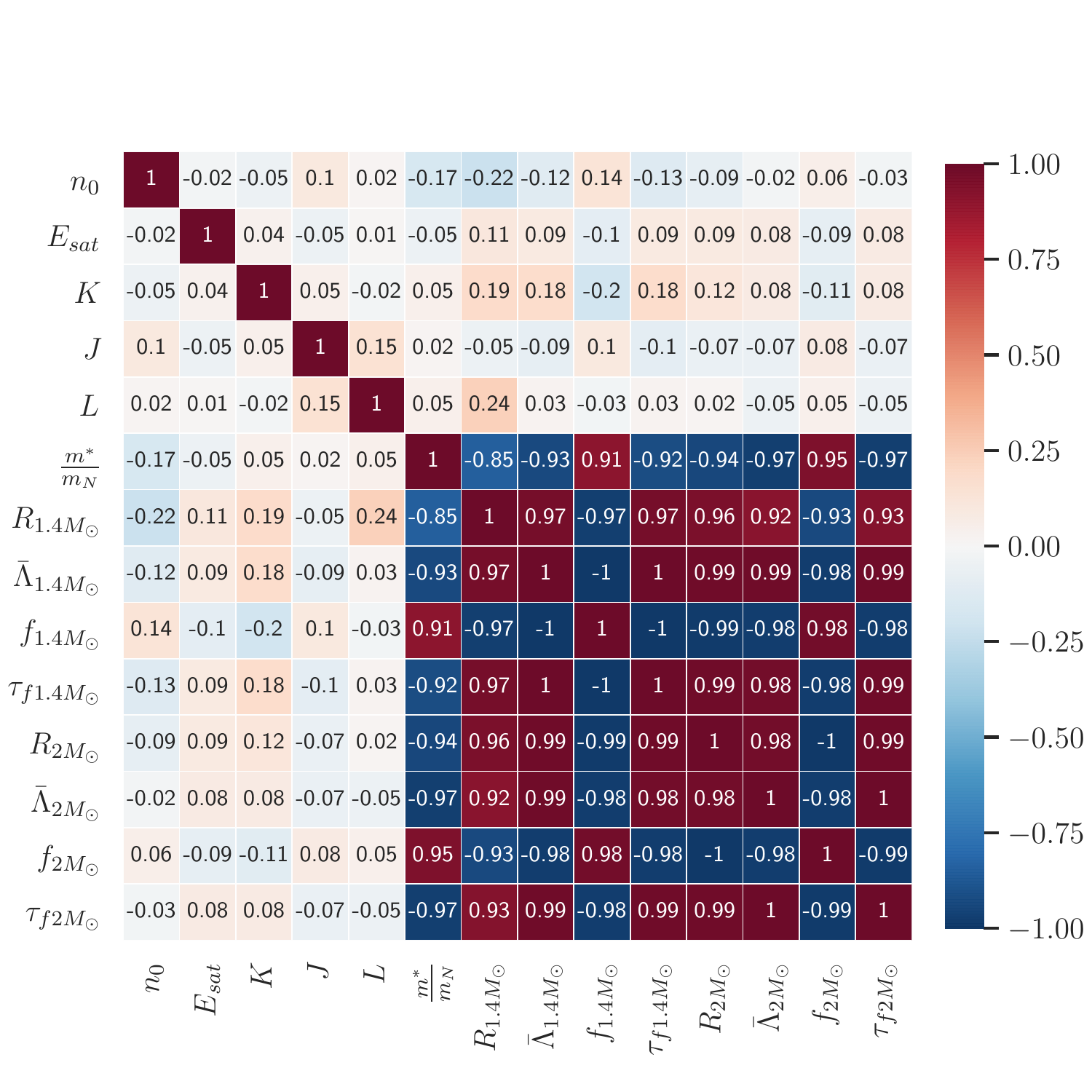}
    \caption{Correlation matrix for nuclear saturation parameters, NS observables and QNM characteristics considering models with $npe\mu$ matter after applying astrophysical constraint ($M_{max}\geq 2M_{\odot}$ and $\bar{\Lambda}_{1.4M_{\odot}}\leq720$) }
    \label{fig:correlation_withnucleon}
\end{figure}

\begin{figure}[htbp]
    \centering
    \includegraphics[width=\linewidth]{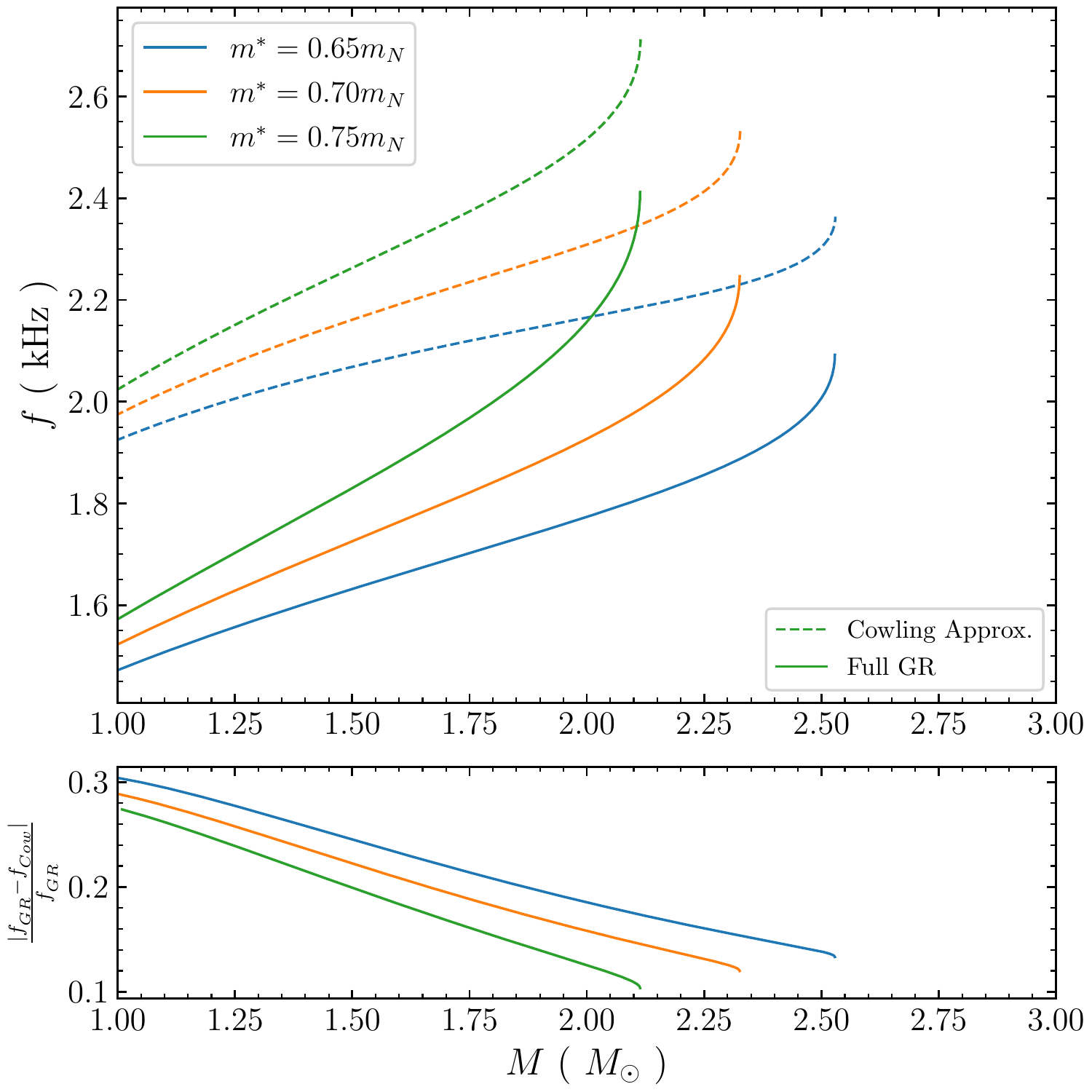}
    \caption{$f$-mode frequencies of stable NS configurations  as a function of stellar mass (upper panel) with $npe\mu$ matter for different nuclear effective mass ($m^*$ at saturation). For comparison, results obtained using Cowling approximation (dashed lines) and obtained in full GR (solid lines) are shown. Lower panel shows relative error for $f$-mode frequencies obtained using two different methods. While varying $m^*$ other parameters are fixed at, $n_0=0.150\ \rm{fm^{-3}},\ E_{sat}=-16.0\ \rm{MeV}, \ J=32\ \rm{MeV}, \ L=60\ \rm{MeV}, \ K=240  \ \rm{MeV} $.}
    \label{fig:fvsm_effm}
\end{figure}

\begin{figure}[htbp]
    \centering
    \includegraphics[width=\linewidth]{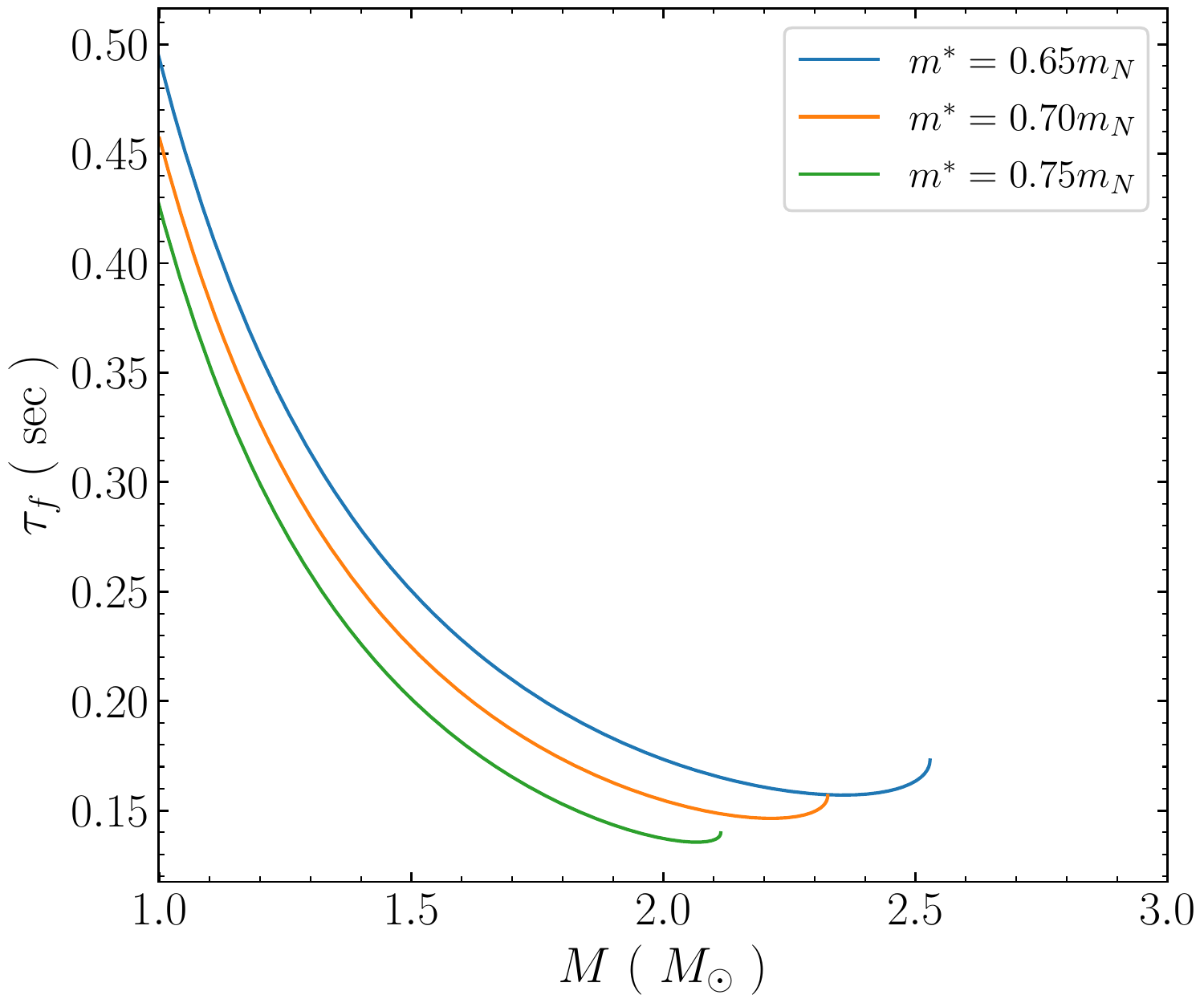}
    \caption{$f$-mode damping time  as a function of stellar mass  with $npe\mu$ matter for different $m^*$. While varying $m^*$ other parameters are fixed to,$n_0=0.150\ \rm{fm^{-3}},\ E_{sat}=-16.0\ \rm{MeV}, \ J=32\ \rm{MeV}, \ L=60\ \rm{MeV}, \ K=240  \ \rm{MeV} $..}
    \label{fig:tauvsm_effm}
\end{figure}

We first consider only nucleonic EoSs to find the effect of nuclear saturation data on NS observables and then extend to involve hyperons. For better understanding, we obtained the Pearson's correlation coefficients ($R_{X_1X_2}$)  among the saturation parameters, NS observables such as radius and tidal deformability and QNM characteristic for canonical $1.4M_{\odot} \text{, and massive } 2M_{\odot} $ NSs. Pearson's linear correlation coefficient ($R_{X_1X_1}$) between two random variables $X$ and $Y$ can be defined as~\cite{ref1}, 
\begin{equation}
    R_{X_1X_2}=\frac{Cov(X_1,X_2)}{S(X_1)S(X_2)}~,
    \label{eq:correlation} 
\end{equation}
where $Cov(X_1,X_2)$ is the co-variance and $S(X_i)$ denotes standard deviation of variable $X_i$. We present the correlation matrix in \Cref{fig:correlation_withnucleon}.   From \Cref{fig:correlation_withnucleon} the following conclusions can be drawn,
\begin{itemize}
    \item NS observables show strong correlation among themselves as well as with the QNM characterstics. As expected from \eqref{eq:tidal_deformability}  $\bar{\Lambda}$ shows a strong correlation  with $R$ (0.97 for $1.4M_{\odot}$ and 0.98 for $2M_{\odot}$). Frequency and damping time also show a strong correlation among themselves.
    \item We find strong correlations  between $f$-mode frequency  and radius which can be explained by looking at \eqref{eq:freqfit} given in \Cref{sec:universal_relations}, similarly the high correlation between  damping time and radius can be explained by \eqref{eq:dampingtimefit} from  \Cref{sec:universal_relations}.
    \item Among the saturation parameters $m^*$ shows strong correlations with radius and tidal deformability ( 0.85 with $R$ and 0.93 with $\bar{\Lambda}$ for $1.4M_{\odot}$) which is expected as $m^*$ is the dominant parameter controlling the stiffness of EoS and hence the radius.
    \item Strong correlations exist between mode characteristics ($f$ and $\tau_f$ ) and nucleon effective mass ($m^*$) for $1.4M_{\odot}$ ( 0.91 with $f$ and 0.92 with $\tau_f$ ) as well as  for $2M_{\odot}$ (0.95 with $f$ and 0.97 with $\tau_f$). This leads us to conclude that the nucleon effective mass has the most dominant effect on the QNM characteristics compared to other nuclear saturation parameters.
\end{itemize}
We display the dependence of frequency and damping time  as a function of stellar mass for variation of $m^*$ in \Cref{fig:fvsm_effm} and \Cref{fig:tauvsm_effm} respectively. 
Extension of our previous calculations from Cowling to involve linearized gravity provides us an opportunity to compare the $f$-mode frequencies from the two different methods. We display a comparison of frequency obtained by two different methods with variation of nucleon effective mass $m^*$ in \Cref{fig:fvsm_effm}. We find Cowling approximation can include  error of  10-30\%  in the quadrupole $f$-mode frequencies and the error decreases with increasing mass. The obtained trend of decreasing error with increasing
mass is in good agreement with the previous result from~\cite{Chirenti2015}. A possible explanation for this trend was discussed in~\cite{Yoshida_2}, given that the $f$-mode eigenfunction is peaked near the surface, increasing mass (or compactness for the given mass range and models considered) can make the metric perturbations less relevant for $f$-mode eigenfunction resulting in a smaller error compared to the frequency obtained within the relativistic Cowling approximation.

\subsubsection{Inclusion of hyperons}

\begin{figure}[htbp]
    \centering
    \includegraphics[width=\linewidth]{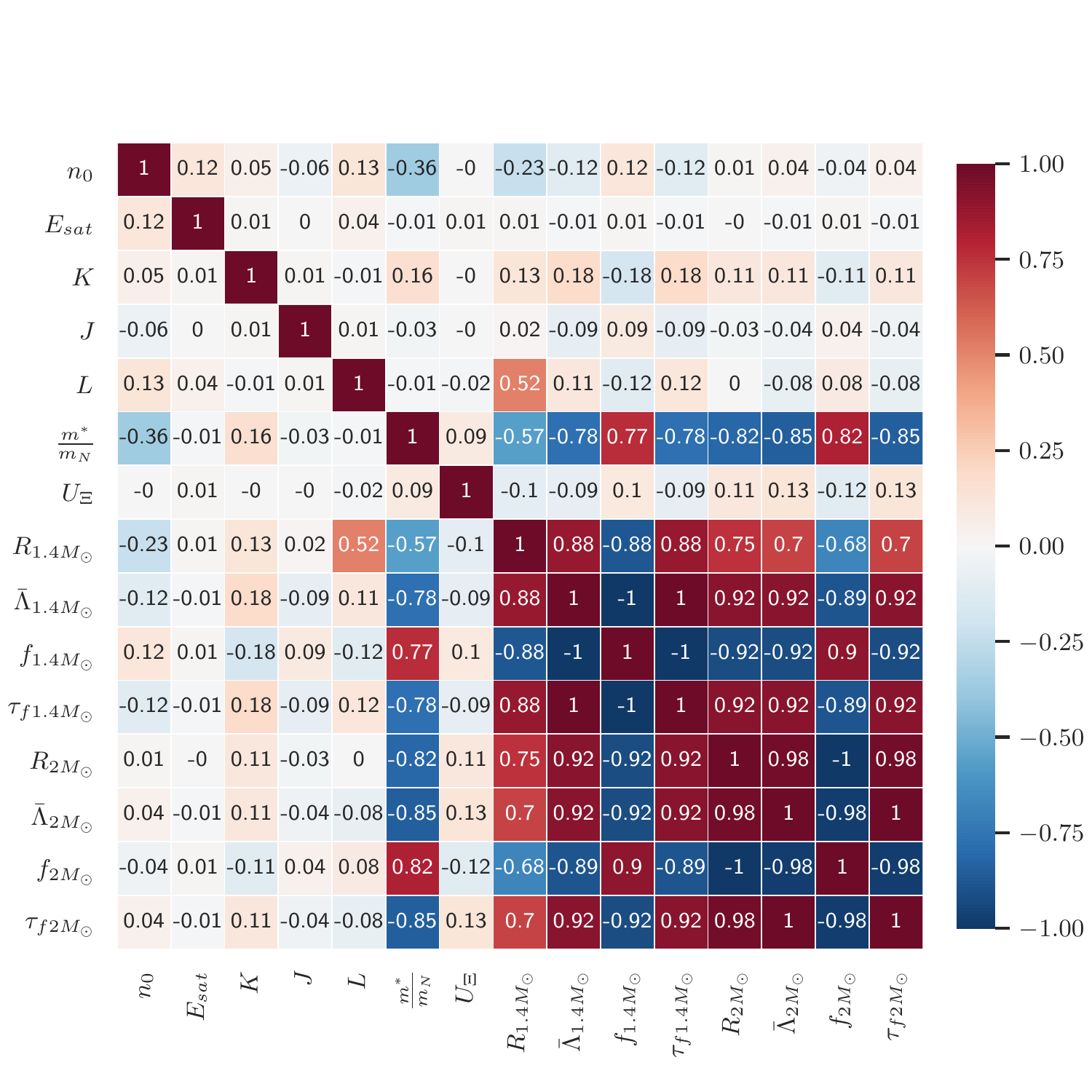}
    \caption{Correlation matrix for nuclear saturation parameters, NS observables and QNM characteristics considering models with $npe\mu Y$ matter after applying astrophysical constraints ($M_{max}\geq 2M_{\odot}$ and $\bar{\Lambda}_{1.4M_{\odot}}\leq720$) }
    \label{fig:correlation_with_ny_matter_2.0}
\end{figure}

We extend our investigation by including presence of hyperons in the NS core and present the correlation matrix in \Cref{fig:correlation_with_ny_matter_2.0} in a  similar fashion as given for nucleonic models. Looking at \Cref{fig:correlation_with_ny_matter_2.0} one can conclude the following:
\begin{itemize}
    \item NS observables show strong correlations among themselves as well as with the QNM characteristics. 
    \item Interestingly the correlation between $L$ and radius of $1.4M_{\odot}$ star increases when compared to the nucleonic case (from 0.24 to 0.52) while the correlation between $m^*$ and $R_{1.4M_{\odot}}$ decreases from 0.85 to 0.57 compared to the nucleonic case.  
    \item However, when it comes to QNM characteristics, they show strong correlations with $m^*$ for a $1.4M_{\odot}$ star as well as for a $2M_{\odot}$ star. It is worth noting that the correlations among $m^*$ and NS observables have decreased compared to the models with only nucleonic EoSs.
\end{itemize}

In light of these inferences from the correlation plot \Cref{fig:correlation_with_ny_matter_2.0}, one can conclude that the $m^*$ has the most dominant effect on the QNM characteristics compared to other nuclear and hypernuclear saturation parameters, even in the presence of hyperons.

\section{Discussions}
\label{sec:discussions}

\subsection{Summary of this work}
We revisit the NS asteroseismology problem in this work, considering realistic EoSs in the  RMF framework and current astrophysical constraints. In a recent publication~\cite{Pradhan2021}, we studied the effect of the influence of the uncertainties in the underlying nuclear and hypernuclear physics on $f$-mode frequencies within the relativistic Cowling approximation. Here, we extend this work to study the effect of uncertainties in the underlying nuclear and hypernuclear physics on complex $f$-mode characteristics (frequency and damping time) of non-rotating perturbed NSs by solving the perturbation equations based on complete linearized equations in general relativity. We provide the asteroseismology relations by considering EoSs with the nuclear and hypernuclear matter in the NS core.
\\

Previous works on NS asteroseismology  or inverse asteroseismology  involved selected realistic or polytropic EoSs. Many of the chosen EoSs have now been rendered incompatible with large NS mass observations or the tidal deformability constraint from merger event GW170817. Hence the empirical relations from past works need to be modified. There are a few efforts to investigate the effect on $f$-modes of the inclusion of exotic forms of matter (hyperon or quark matter) ~\cite{Salcedo, Benhar,Sandoval,Vasquez,kumar2021,das2021, Flores2017} or to improve the asteroseismology relations with current astrophysical constraints ~\cite{shashank2021,Sotani_Bharat2021,Zhao2022b,Mu2022}. However, the works are either limited to selective EoSs or used Cowling approximation to find the mode characteristics. 
\\

As mentioned, the extension of our previous work~\cite{Pradhan2021} (where relativistic Cowling approximation was used to find the mode frequency) by involving complete linearized equations of general relativity, allows us to compare the mode frequencies obtained within the two different methods. For the models considered, frequencies obtained using relativistic Cowling approximation can include an error up to 30\% compared to those obtained in full general relativity. Solving the NS oscillation in full general relativity enables us to investigate the effect of uncertainties in the underlying nuclear and hypernuclear physics on both frequency and damping time of the $f$-mode. The dependence of mode frequencies on the stellar mass remains qualitatively similar (increasing) to results obtained using relativistic Cowling approximation irrespective of the composition of the NS interior. $f$-mode damping time shows an inverse relation with NS mass for both nuclear and hypernuclear matter EoSs. Considering NS masses starting from $1M_{\odot}$ and  up to the possible maximum stable NS mass configuration of each model along with current astrophysical constraints, the frequency and damping time of quadrupole $f$-mode oscillations are found to be in the range of  $1.47\ \rm{kHz}-2.45$kHz and 0.13 sec $-$ 0.51 sec respectively. In this work, we also obtained the UR involving tidal deformability and mass scaled $f$-mode characteristics. We tested the hypotheses of universality between tidal deformability for a canonical $1.4M_{\odot}$ NS. Using the upper bound of tidal deformability, we found the lower bound on the $f_{1.4M_{\odot}}$ to be 1.6kHz, which is in agreement with the result obtained with Bayesian estimation from Pratten et~al.~\cite{Pratten2020}. We also found that, using the upper bound of the tidal deformability coming from the event GW170817 for a $1.4M_{\odot}$ ( i.e., $\bar{\Lambda}\leq720$ ), the damping time of a $1.4M_{\odot}$ has an upper limit of 0.28 sec. From ~\Cref{fig:f_tidal_1.4m}, it is evident that frequencies above 1.7 kHz are difficult to describe with hyperonic EoSs in our model.
\\

We found that while considering nucleonic EoSs and imposing current astrophysical constraints, among all the nuclear saturation parameters, the nuclear effective mass $m^*$ has the most dominant effect on $f$-mode characteristics. We explored correlations among saturation parameters, NS observables (radius and tidal deformability) and $f$-mode characteristics for a canonical $1.4M_{\odot}$ and a massive $2M_{\odot}$. NS observables show strong correlations among themselves as well as with $f$-mode characteristics. 
The strong correlations of $m^*$ with NS observables ($R_{1.4M_{\odot}}$, $\bar{\Lambda}_{1.4M_{\odot}}$) and $f$-mode characteristics ($f_{1.4M_{\odot}}$, $\tau_{f1.4M_{\odot}}$) for NS with $1.4M_{\odot}$ remain so even for $2M_{\odot}$. 
\\

We further investigate the effect of uncertainties in nuclear and hypernuclear saturation parameters on the $f$-mode characteristics by considering the presence of hyperons on the NS interior along with the imposition of current astrophysical constraints. 
We checked that even in the presence of hyperons, the nuclear effective mass ($m^*$) still has the dominant effect on the $f$-mode characteristics. The hypernuclear parameter $U_{\Xi}$ was found to have a minor effect on $f$-mode characteristics. 
We also provide correlations among the nuclear saturation parameters, hypernuclear parameters as well as with the NS observables (radius and tidal deformability) and $f$-mode characteristics for a $1.4M_{\odot}$ and a massive $2M_{\odot}$ for hyperonic stars. Similar to nuclear-matter models, NS observables show strong correlations among themselves and also with $f$-mode characteristics. Considering the presence of hyperons in the NS core and imposing the maximum mass limit of $2M_{\odot}$  and $\bar{\Lambda}_{1.4M_{\odot}}\leq720$, the correlation between slope of symmetry energy at saturation ($L$) and $R_{1.4M_{\odot}}$ increases in comparison with nucleonic models whereas the correlation between $m^*$ and  $R_{1.4M_{\odot}}$ decreases.  
\\

In our analysis, we provide URs in asteroseismology considering the entire parameter range of uncertainties within the framework of the RMF model compatible with state-of-the-art nuclear and hypernuclear physics subject to current astrophysical constraints. These empirical relations involving $f$-mode frequency and average density or appropriately scaled damping time and stellar compactness differ in their fit parameters compared to those proposed previously in the literature~\cite{Andersson96, Andersson98, Benhar}. 
Also in full general relativity, we found the empirical fits between frequency and density to be dependent on the choice of EoSs considered. We then tested the hypotheses of universality between stellar compactness and $f$-mode characteristics scaled with stellar mass. We provide a quadratic universal relation among mass scaled angular frequency and stellar compactness (Re($M\omega$)=$F_1(M/R)$) and a universal relation for mass scaled damping time and stellar compactness (i.e.,Im($M\omega$)=Im($M/\tau_f$)=$F_2(M/R)$). When the angular frequencies and damping times are scaled appropriately with NS mass, the universality between $f$-mode frequency and damping time can be described by the proposed UR. 

\subsection{Future prospects}
During the inspiral phase of a binary NS merger,  when the tidal field reaches resonance with the NS internal oscillation modes, particular new features are created in the GW  waveform that if detected can provide information on the QNMs. Amongst these modes, the $f$-mode is the most important one. Hence studying the effect of dynamical tides can be used to analyze the influence of QNM modes in the merger waveform~\cite{Schmidt2019, Pratten2020}. In merger events, tidal deformability is another important observable parameter. Future detection will put tight constraints on tidal deformability;  hence universal relations involving $f$-mode characteristics and tidal deformability are useful for analyzing the mode characteristics.  
\\
 
To understand the $f$-modes thoroughly, other complicating effects such as rotation ~\cite{Doneva, Steinhoff2021,Kruger2021}, magnetic fields, the effect of superfluidity \cite{Gualtieri2014}, and the presence of deconfined quark core should be taken into account. Superfluidity will play a role in the case of cold NSs, whereas rotation will play a crucial role in the hot and differentially rotating merger remnant. It has been shown that for stars with deconfined quarks in NS interior, the $f$-mode vs tidal deformability relation deviates from the universality in isolated NSs~\cite{Wen} as well as in merger remnants~\cite{Blacker2020}. 

\subsection{Detectability}
We conclude by making some remarks on the detectability of the $f$-mode of hyperonic stars. As the $f$-mode amplitude peaks near the star's surface, it may be excited strongly by glitching behaviour in an isolated star, or by tidal forces due to a companion during the late inspiral in a merger. In the former case, one would expect a GW burst in the detector, while in the latter case, the $f$-mode would draw energy from the orbit, affecting the phase of the gravitational waveform. The study by Pratten et al.~\cite{Pratten2020} placed a lower bound on the quadrupolar $f_2$-mode in the region of 1400 Hz, including the mode excitation directly as a parameter in the analysis of data from GW170817. Here, we will consider isolated hyperonic stars that emit a burst of GW due to the $f$-mode and estimate the peak gravitational wave strain and associated energy required for detection in aLIGO and third generation detectors. Utilizing the methodology in~\cite{PhysRevD.101.103009}, wherein the burst waveform is modelled as an exponentially damped oscillation with frequency $\nu_f$ and damping time $\tau_f$, we find the peak strain

\begin{equation}
h_{0}=1.46 \times 10^{-13} \sqrt{\frac{E_{\rm gw}}{M_{\odot}c^2}} \sqrt{\frac{1 {\rm sec}}{\tau_f}} \frac{1 \rm{kpc}}{d}\left(\frac{1 \rm{kHz}}{\nu_f}\right)~.
\end{equation}

\begin{figure}[ht]
    \centering
    \includegraphics[width=\linewidth]{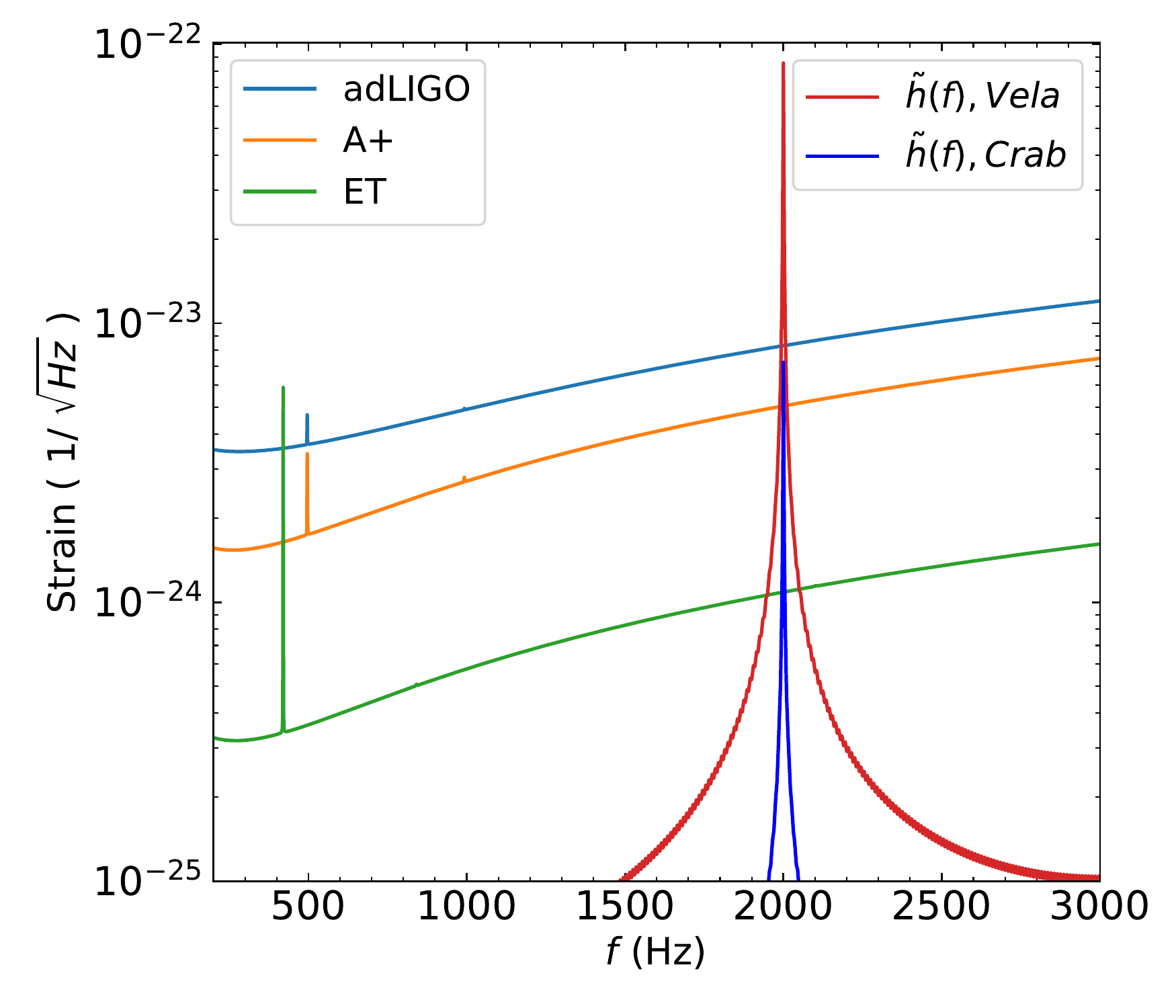}
    \includegraphics[width=\linewidth]{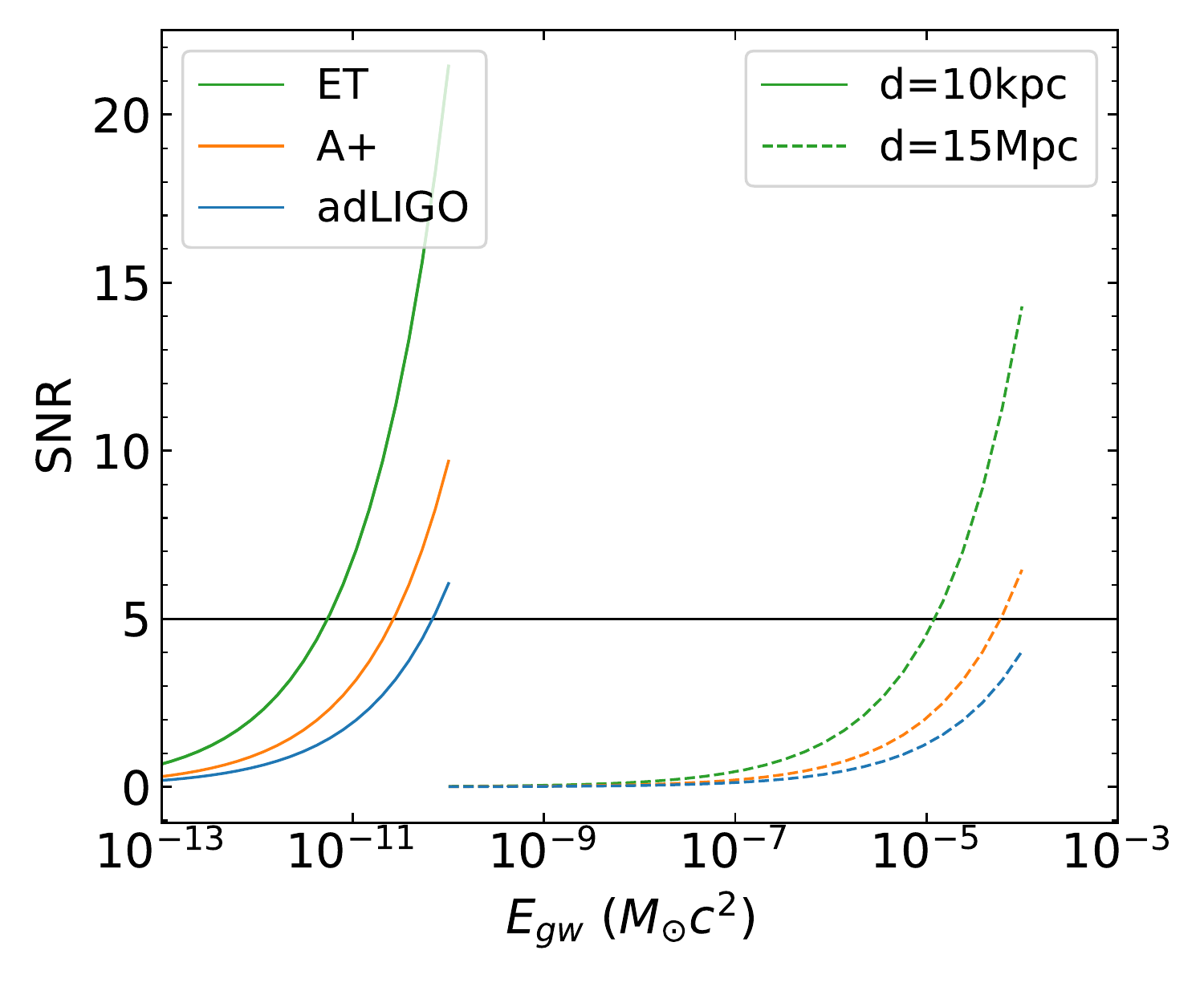}
    \caption{Top panel: Gravitational wave strain corresponding to a maximum glitch energy observed in Vela pulsar (Crab pulsar) with distance to be 290 pc (2 kpc) are shown along with the sensitivity bands of different configurations. The waveforms are generated using the GW inference package Bilby~\cite{Bilby_2019}. (lower panel) : Signal to noise ratio (SNR) for a source with f-mode frequency $1.70$kHz and damping time $0.25$ sec at different detector configurations as a function of $E_{\rm gw}$ for sources to be at 10kpc (solid lines) or at 15Mpc (dashed lines).}
    \label{fig:strain_and_snr}
\end{figure}

Choosing a canonical NS mass of 1.4$M_{\odot}$ at a distance of 10 kpc, an $f$-mode frequency of 1.70 kHz with a damping time of 0.25 sec, and assuming that $E_{\rm gw}$ is of the order of a glitch in the Vela pulsar and highly efficient in producing GW, \Cref{fig:strain_and_snr} (top panel) shows the resulting frequency domain waveform   $\tilde{h}( f)$ against the sensitivity curve of advanced LIGO (adLIGO) ~\cite{KAGRA:2013rdx}, A+ ~\footnote{\url{https://dcc.ligo.org/LIGO-T2000012/public}} and the Einstein Telescope (ET) ~\cite{Hild_2011}\footnote{\url{https://dcc.ligo.org/LIGO-T1500293/public}} and the $E_{\rm gw}$ (bottom panel) required for the typical signal-to-noise ratio (SNR$\approx$ 5) for detection in these instruments.
For sources at 10 kpc  for signal to noise ratio (SNR) $\geq 5$, the energy $E_{gw}$ should be greater than $5.75\times 10^{-12} M_{\odot}c^2,\ 2.88\times 10^{-11} M_{\odot}c^2,\ \text{and } 6.88\times 10^{-11} $ for ET, A+ and adLIGO configuration respectively.  For sources at 15 Mpc for SNR $\geq 5$ in A+ and ET configuration $E_{gw}$ should be greater than $6.55\times 10^{-5} M_{\odot}c^2,\ \text{and}\  1.3\times 10^{-5} M_{\odot}c^2 $ respectively. As $E_{\rm GW}$ is a parameter in the waveform, depending upon the distance, GWs induced from (i) NS glitches (ii) a supernova explosion (iii) a prominent phase transition, leading to a mini collapse in NS  are possible sources that could be detectable. These might be considered optimistic estimates, given that most glitches are weaker than in Vela pulsar and might not couple that strongly to GW. In addition, individual sources may have considerable uncertainty in either the distance or radius parameters. The statistical approach followed in~\cite{PhysRevD.101.103009} uses the BSk24 EOS (no hyperons or exotic matter) to model ordinary neutron stars satisfying current observational constraints, and suggests that third generation detectors like ET and Cosmic Explorer (CE) would offer the best chance to detect the transient bursts from $f$-modes. 

\section{Acknowledgements}
\label{sec:ack}
P.J. is supported by the U.S. National Science Foundation Grant No. PHY-1913693. B.-K.P. and D.C. acknowledge usage of the IUCAA HPC computing facility for the numerical calculations. B.-K.P. is thankful to Suprovo Ghosh, Dhruv Pathak, Swarnim Shirke and Tathagata Ghosh for their useful discussions during this work.

\newpage

\bibliography{Pradhan_Resub}

\end{document}